\documentclass[12pt]{elsarticle}
\pdfoutput=1
\usepackage{setspace}
\usepackage{amsmath}
\usepackage{booktabs}
\usepackage{graphicx,subcaption}
\usepackage{listings}
\graphicspath{ {images/} }
\usepackage{float}
\usepackage{caption}

\usepackage{array}
\usepackage{geometry}
\usepackage[font=small,labelfont=bf]{caption}
\makeatletter\def\CT{\def\@captype{figure}}\makeatother
\usepackage{pifont}
\usepackage{array}
\usepackage{ctable}
\usepackage[titletoc,title]{appendix}
\numberwithin{equation}{section} 
\usepackage{hyperref}
\hypersetup{
    colorlinks=true,
    linkcolor=blue,
    filecolor=magenta,      
    urlcolor=blue,
}
 
\makeatletter
\def\ps@pprintTitle{%
   \let\@oddhead\@empty
   \let\@evenhead\@empty
   \def\@oddfoot{\reset@font\hfil\thepage\hfil}
   \let\@evenfoot\@oddfoot
}
\makeatother

\begin{document}

\begin{frontmatter}
\title{Quantifying the Trade-offs between Energy Consumption and Salt Removal Rate in Membrane-free Cation Intercalation Desalination}

\author[add1]{Sizhe Liu}
\ead{sliu135@illinois.edu}

\author[add1,add2,add3,add4]{Kyle C. Smith\corref{cor1}}
\ead{kcsmith@illinois.edu}

\cortext[cor1]{corresponding author}
\address[add1]{Department of Mechanical Science and Engineering, University of Illinois at Urbana-Champaign, Urbana, IL 61801, USA.}
\address[add2]{Department of Materials Science and Engineering, University of Illinois at Urbana-Champaign, Urbana, IL 61801, USA.}
\address[add3]{Computational Science and Engineering Program, University of Illinois at Urbana-Champaign, Urbana, IL 61801, USA.}
\address[add4]{Beckman Institute for Advanced Study, University of Illinois at Urbana-Champaign, Urbana, IL 61801, USA.}

\begin{abstract}
Electrochemical desalination devices that use redox-active cation intercalation electrodes show promise for desalination of salt-rich water resources with high water recovery and low energy consumption. While previous modeling and experiments used ion-exchange membranes to maximize charge efﬁ-ciency, here a membrane-free alternative is evaluated to reduce capital cost by using a porous diaphragm to separate Na$_{1+x}$NiFe(CN)$_6$ electrodes.  Two-dimensional porous-electrode modeling shows that, while charge efficiency losses are inherent to a diaphragm-based architecture, charge efficiency values approaching the anion transference number (61$\%$ for NaCl) are achievable for diaphragms with sufficiently low salt conductance.  Closed-form equations are thereby derived that relate charge efficiency to the non-dimensional P\`{e}clet and Damk{\"o}hler numbers that enable the selection of current and flow velocity to produce a desired degree-of-desalination.  Simulations using these conditions are used to quantify the tradeoffs between energy consumption and salt removal rate for diaphragm-based cells operated at a range of currents.   The simulated distributions of reactions are shown to result from the local salt concentration variations within electrodes using diffusion-potential theory.  We also simulate the cycling dynamics of various flow configurations and show that flow-through electrodes exceed the degree-of-desalination compared with flow-by and flow-behind configurations due to solution stagnation within electrodes.

\end{abstract}
\end{frontmatter}

\section{Introduction}

Global water shortage has led to energy intensive water extraction and treatment methods including desalination \cite{almarzooqi2014application}. Further, water resources are depleting due to exogenous factors, including pollution \cite{mahar2001optimal} and salt intrusion in groundwater due to seawater-level rise \cite{SefGW2014,EssWR2010,Dasgupta2014}.  A variety of desalination technologies have been developed during the past 50 years, including pressure-driven reverse osmosis, thermally driven multi-stage flash and multi-effect distillation, electrodialysis (ED), and capacitive deionization (CDI) \cite{Elim2016DesalMater,khawaji2008advances,van2002distillation,Suss2015Review}.  
CDI technologies conventionally use electric ﬁeld as a driving force to absorb ions into electric double-layers (EDLs) and have shown promise for low energy consumption in desalinating brackish waters \cite{Zhao2013,Suss2015Review}.  Salt adsorption capacity for such EDL-based CDI devices is limited by polarizable surface area of electrodes \cite{Suss2015Review}.  To extend the limits of salt removal beyond EDL-based CDI Smith and co-workers \cite{smith2016ion,smith2017theoretical} introduced a novel desalination device concept using electrodes containing redox-active intercalation host compounds (IHCs), materials commonly used in aqueous rechargeable batteries \cite{Kim2014,Whitacre2010,Wessells2011,Li2013}, that adsorb cations in the bulk of the host lattice upon reduction of electroactive species in the IHC, including Na$_{0.44}$MnO$_2$ \cite{smith2016ion, smith2017theoretical}, NaTi$_2$(PO$_4$)$_3$ \cite{smith2016ion}, and NaNiFe(CN)$_6$ \cite{smith2017theoretical}.  Inspired by the salt depletion/accumulation processes that are known to occur within Li-ion batteries \cite{Doyle1997}, the so-called cation intercalation desalination (CID) process was shown to desalinate in one electrode, while simultaneously concentrating salt in the other \cite{smith2016ion}.

Initial predictions of CID \cite{smith2016ion} showed that a cation-blocking ion-exchange membrane (IEM) was needed to approach 100$\%$ charge efficiency (defined as the ratio salt moles removed to the moles of electrons transferred), while a thin battery-type separator produced only 35$\%$ charge efficiency.  Following from these findings the CID concept was demonstrated experimentally using IEMs in brackish water with identical electrodes \cite{Porada2017} and in seawater with dissimilar electrodes \cite{Lee2017}.  Furthermore, later predictions showed that a series of IEMs with alternating selectivity can be stacked between IHC electrodes to increase water salt adsorption capacity \cite{smith2017theoretical}, which was subsequently demonstrated experimentally \cite{Kim2017}.  Similarly, in membrane CDI (MCDI) IEMs are used in conjunction with EDL-based electrodes to enhance charge efficiency \cite{LEE2006125, ZhaoEES2012} by preventing the charge-efficiency losses incurred in CDI without IEMs due to co-ion repulsion \cite{BIESHEUVEL20151} and salt residual within electrodes \cite{shang2017combined}.  From an economic standpoint charge efficiency losses impact the operating costs arising from desalination energy consumption.  These operating costs are inversely proportional to the energy-normalized salt adsorption (ENAS) in CDI.  Despite the potential to reduce operating cost using IEMs, they impart substantial capital cost \cite{DarlingEES2014} that decreases with increasing average salt adsorption rate (ASAR) in CDI.  Because ENAS and ASAR result from design and operating conditions (and are not determined independently), the tradeoffs between them have recently been explored in both the CDI \cite{Hawks2018} and MCDI \cite{Kim2014,Wang2018} contexts.  Recently, a contingent of CDI researchers communicated that intercalation-based desalination devices belong to the CDI technology class because their performance can be analyzed using similar metrics \cite{biesheuvel2017capacitive}.  While the electrosorption mechanism of EDL-based CDI electrodes and the cell architectures used to harness them differ substantially from CID, we use these metrics presently to quantify energy consumption and salt adsorption when using CID cells without IEMs.  

Aside from the cost of IEMs, their permselectivity is imperfect and their resistance is coupled to it \cite{geise2013ionic, geise2013salt}, while also possessing a finite lifespan \cite{strathmann2010electrodialysis,shin2013review,cui1998development,la2011batteries}.  Though the durability of IEMs has been enhanced by using nanomaterials \cite{kou2014graphyne,corry2008designing,dzyazko2017composite}, the production of economical, long-lasting IEMs with high electrical conductivity is still a challenge.  While there is substantial room to optimize the performance of CID using IEMs, in the broader context of water treatment opportunities it is also prudent to consider alternatives to IEM-based CID cells that require minimal capital investment with a marginal increase in energy consumption.

Accordingly in this work, we use porous electrode modeling to study the cycling behavior of CID cells that use diaphragms instead of IEMs.  Presently we use a membrane-free approach with two electrodes that only adsorb cations, in contrast with other approaches where one electrode adsorbs cations and the other anions \cite{pasta2012desalination,Nam2017,Chen2017,Srimuk2016}.  Here, we define a diaphragm as an inert, non-conducting porous layer that is imbibed with electrolyte, that possesses low hydraulic permeability, and adjacent to which solution flows.  Since the 1920s such diaphragms have been used in brine electrolysis (i.e., chlor-alkali processing) to produce chlorine gas and caustic soda simultaneously \cite{crook2016chlor}.  Early diaphragms were comprised of paper, and later asbestos, to reduce caustic-soda transport between electrodes \cite{hass1976developments}. While the use of asbestos in diaphragms has driven the chlor-alkali industry toward the use of IEMs instead of diaphragms, environmental-friendly alternatives have been invented recently \cite{d1984separator,broman2002redox} and novel diaphragm materials are also conceivable.  Non-selective filters have also been used to separate cation intercalation electrodes in salinity gradient energy harvesting devices \cite{Logan2016SalGrad}. Furthermore, non-selective separators \cite{yang2013nanostructured,yang2013membrane}, flow configurations \cite{braff2013membrane,suss2016membraneless}, and macromolecular redox-active units \cite{Nagarjuna2014} have recently been used in redox flow batteries (RFBs) to eliminate the use of IEMs.  In particular, in the context of laminar RFBs understanding the balance between flow rate and mass transfer processes is essential to minimizing crossover \cite{braff2013membrane,suss2016membraneless}. 

In this work, we simulate the trade-offs between energy consumption and salt removal rate for a membrane-free cation intercalation desalination device that uses a porous diaphragm to limit salt diffusion between electrodes. To illustrate the cycling dynamics of such a device we consider electrodes comprised of Na$_{1+x}$NiFe(CN)$_6$ (nickel hexacyanoferrate) cation intercalation material using a regular solution model for equilibrium potential with solution flowing through electrodes. We explore the effect of the diaphragm's design and operating current density from which a reduced-order analytical model and charge-efficiency limits are ultimately established. We use this analytical model to explore operating conditions that span a range of the ENAS/ASAR space, while maintaining a tight range of degree-of-desalination.  With these operating conditions we simulate the two-dimensional charge transport processes within electrodes and the diaphragm, and we compare the cell's performance with and without electrolyte recycling and using alternative flow configurations adjacent to electrodes, rather than through them.

\section{Methodology}
\subsection{System definition and materials}

\begin{figure}[H]
    \centering
    \includegraphics[width=\textwidth]{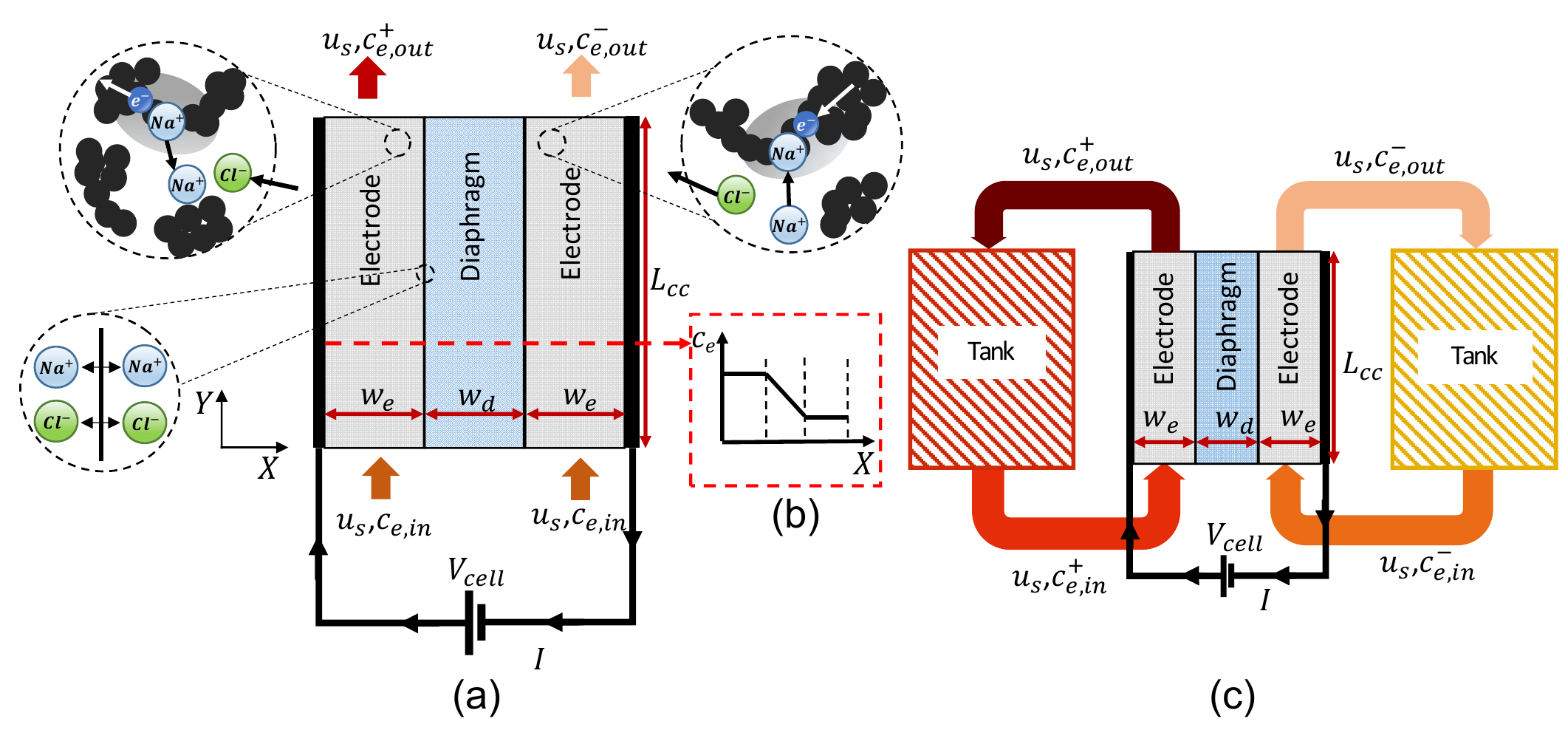}

    \caption{Schematic of a diaphragm-based cation-intercalation desalination device (a-b) without and (c) with effluent recycling.  In (a) Na$^+$ ions in the positive electrode (left) de-intercalate from IHC into the electrolyte, while Na$^+$ ions in the negative electrode (right) intercalate into IHC. Both Na$^+$ and Cl$^-$ ions migrate and diffuse through the diaphragm. (b) A simplified concentration distribution along the $X$-direction is later assumed to derive a separate analytical model.  (c) The recycling cell has two tanks attached to the reactor to enable the recirculation of effluent.}
    \label{fig:cell schematic}
\end{figure}

We simulate the time-dependent dynamics of a two-dimensional CID device using two cation-intercalating porous electrodes of thickness $w_e$ and porosity $\epsilon_e$ separated by a diaphragm of thickness $w_d$ and porosity $\epsilon_d$, as shown in Fig.~\ref{fig:cell schematic}.  On the current collector a fixed current density $i$ is applied over its length $L_{cc}$.  We model the diaphragm as an uncharged porous layer with finite thickness, in contrast with our previous work with charged IEMs of infinitesimal thickness \cite{smith2016ion,smith2017theoretical}.  For the majority of this study flow through (FT) the respective electrodes is assumed, and flow through the diaphragm is neglected in all instances.  This assumption is valid for sufficient contrast between the hydraulic permeability $k$ of the electrodes and the diaphragm.  For the FT electrode configuration we model the flow of aqueous NaCl solution with a uniform superficial velocity $\vec u_s = u_s \hat{j}$ within the porous electrodes.  We note that the FT configuration has only been modeled \cite{smith2016ion, smith2017theoretical} and has yet to be implemented experimentally in CID, but similar configurations are used readily in redox flow battery (RFB) reactors \cite{darling2014influence} and their development in CID is a subject of our on-going research.  Other flow configurations are possible, including flow-behind \cite{Porada2017} and flow-by \cite{smith2017theoretical} electrodes.  Accordingly, we also compare the cycling dynamics between these three flow configurations.  For these cases we use fully developed parabolic velocity profiles within open flow channels with a thickness $w_c$ chosen to match the electrode solution volume used in FT simulations, as in Ref.~\cite{smith2017theoretical}.
For all flow configurations the salt concentration in solution $c_e$ deviates from the inlet concentration $c_{e,in}$ to $c_{e,out}^+$ and $c_{e,out}^-$ within the positive and negative electrode effluent, respectively.

In both electrodes we model the intercalation of Na$^+$ ions into and out of nickel hexacyanoferrate (NiHCF), a type of Prussian Blue analogue with facile kinetics \cite{Mizuno2013}, an ability to intercalate a variety of cations \cite{WessellsJES2011}, and belonging to a class of materials that have recently been predicted \cite{smith2017theoretical} and demonstrated \cite{Porada2017,Kim2017,Lee2017} in CID devices using IEMs.  Here Na$^+$ ions intercalate into NiHCF when it is electrochemically reduced:
\begin{equation}
    x_{Na}Na^+ + x_{Na}e^- + NaNiFe(CN)_6 \rightarrow Na_{1+x_{Na}}NiFe(CN)_6,
\end{equation}

\noindent where $x_{Na}$ is the fraction of intercalated Na$^+$ within the IHC that spans between zero and unity.  We initialize the positive and negative electrodes to reduced ($x_{Na}$=99.958$\%$) and oxidized ($x_{Na}$=0.042$\%$) states respectively, enabling a ``symmetric'' cell \cite{smith2016ion} that operates as a reversible rocking-chair battery \cite{Scroscatti1995} when charged and subsequently discharged.  The operating concept for the present cell follows that described originally by Smith and co-workers \cite{smith2016ion}: Na$^+$-rich IHC in the positive electrode releases Na$^+$ ions into electrolyte upon oxidation, while Na$^+$-deficient IHC in the negative electrode absorbs Na$^+$ ions from electrolyte upon reduction.  These processes induce a salt concentration gradient through the diaphragm, and both Na$^+$ and Cl$^-$ ions transport between the electrodes to facilitate current in the cell.  As a result, effluent streams of two different concentrations are generated (Fig. \ref{fig:cell schematic}).  Because the direction of the applied current is switched during discharging, dilution and concentration processes swap between the respective electrodes, and the containers for the concentrated and diluted effluent are swapped accordingly.  In contrast with the processes described in Refs.~\cite{smith2016ion,smith2017theoretical} using ideally permselective IEMs, the degree of salt dilution and concentration is affected by the rate at which Na$^+$ ions transport through the diaphragm. 

In addition to a cell architecture where effluent directly exits as produced water (Fig.\ref{fig:cell schematic}a), we also simulate a cell architecture wherein effluent is recycled to tanks from which influent is further drawn (Fig.\ref{fig:cell schematic}c).  When such plumbing is implemented, this architecture produces influent salt concentration that varies throughout a complete electrochemical cycle. Therefore, we use $c_{e,in}^{+}$ and $c_{e,in}^{-}$ to specify the influent salt concentration for the respective electrodes. We initialize the tank concentrations at the beginning of the charge and discharge steps to the salinity level of the source water being treated $c^0_e$.  In all simulations we initialize salt concentration in the entire cell to $c^0_e$.   Hereafter we refer to the cell with effluent recycling as a recycling (RC) cell and the cell without effluent recycling as a non-recycling (NRC) cell.  An RC cell design has been employed in previous research as a mean to promote the absorption degree of certain toxic elements in waste water \cite{amor2001fluoride,frateur2007adsorption,menoret2002use}. Further, experiments have shown that effluent recycling in an ED system enhances adaptability in practical treatment scenarios \cite{amor2001fluoride}.

\subsection{Electrochemical model}
A porous-electrode model is used here to simulate solution-phase ion diffusion and migration, electron conduction through porous electrodes, and intercalation reactions at IHC/solution interfaces. We use the theory described in Ref.~\cite{smith2017theoretical}, which incorporates a concentrated-solution description of migration and diffusion within NaCl electrolyte, a regular solution of cations and vacancies for the equilibrium cation intercalation potential of NiHCF, and facile intercalation kinetics for NiHCF.  We note that higher-order models, such as the Temkin equation used in \cite{Porada2017}, can be used to describe the equilibrium potential.  A regular solution model agrees within 40 mV of low C-rate experimental cycling data \cite{smith2017theoretical}. 

Here, we emphasize the ``salt transport equation,'' based upon which we later develop an analytical model used (1) to identify efficient operating rate regimes for CID devices using diaphragms and (2) to correlate those operating conditions through non-dimensional parameters.  We refer the reader to our earlier work for the detailed governing equations \cite{smith2017theoretical,smith2016ion}.  Na$^+$ ions deplete from the electrolyte as Na$^+$ ions intercalate into NiHCF, such that the balance of ionic species in electrolyte preserves solution-phase electroneutrality \cite{smith2016ion}. By homogenizing the individual conservation equations for Na$^+$ and Cl$^-$, a potential-eliminated ``salt transport equation'' can be derived that governs the local evolution of salt concentration $c_e$ \cite{smith2016ion,smith2017theoretical}:

\begin{equation}
    \frac{\partial (\epsilon c_{e})}{\partial t} + \nabla \cdot (\vec u_{s}c_{e})+\nabla \cdot (-\tilde D_{eff}\nabla c_{e}) - a\upsilon_{s}t_{-}\frac{i_n}{F}=0.
    \label{e1}
\end{equation}

 \noindent This equation incorporates the effect of intercalation reactions as a source term proportional to the intercalation reaction current density $i_n$ at the surface of IHC particles, the surface area per unit volume $a$ and volume fraction $\upsilon_s$ of IHC particles, and the transference number of anions within the electrolyte $t_-=1-t_+$, where $t_+$ is the cation transference number.  $F$ is Faraday's constant.  The effective chemical diffusivity of salt $\tilde D_{eff}$ is a fraction of the bulk chemical diffusivity of salt $\tilde D$ as a result of tortuosity and porosity effects that we model here using the Bruggeman approximation $\tilde D_{eff}=\epsilon^{1.5} \tilde D$.  We use experimental data for $\tilde D$ \cite{rard1979mutual} and calculate the bulk ionic conductivity $\kappa$ consistent with concentrated solution theory \cite{lai2011mathematical} using aqueous-NaCl activity coefficients from experiment \cite{Hamer1972}, as in Ref.~\cite{smith2017theoretical}.

 NiHCF particles have small size and high rate capability \cite{wu2013low,wessells2011nickel,Mizuno2013}.  Consequently, we neglect kinetic polarization and solid-state mass-transfer resistance within NiHCF particles, as in our previous work \cite{smith2017theoretical}. In practice, we employ Butler-Volmer kinetics \cite{smith2016ion} by assuming a reaction rate constant of $2\times10^{-11} \text{mol m}^{-2}s^{-1}$ per $(\text{mol m}^{-3})^{1.5}$, the volumetric surface area of 50 nm particles \cite{smith2017theoretical} and a maximum specific capacity of 59 $\text{mAh g}^{-1}$ (based on low rate cycling data in Refs.~\cite{wu2013low,wessells2011nickel}). The terminal concentration of intercalated Na$^+$ ions $c_{s,max}$ was taken as 4045.5 $\text{mol m}m^{-3}$ to normalize concentration of intercalated Na$^+$ ions and thereby calculate the intercalated-Na fraction $x_{Na}$. The porous electrodes simulated in this study include 50 vol.$\%$ NiHCF and 40 vol.$\%$ porosity, the remaining volume being available for inert binder and conductive additives.  An effective electronic conductivity of 100 $\text{S m}^{-1}$ is assumed for the porous electrodes studied, as in our previous work \cite{smith2017theoretical}.  We note that ionic and electronic transport limitations at the scale of NiHCF particles could be important at sufficiently high currents.  Here, we neglect such effects as in Ref.~\cite{smith2017theoretical}. All simulations use common electrode dimensions (1.0 mm electrode thickness $w_e$ and 20 mm channel length $L_{cc}$).

\section{Results and discussion}
The development of CID cells using diaphragms instead of IEMs not only requires judicious choice of IHCs, but also appropriate choice of design and operating parameters, including diaphragm dimensions,  applied current density, and influent flow rate. We illustrate the impact of the first two parameters by varying their values for an NRC cell.  Then, we establish the relationships between these parameters and charge efficiency using dimensionless numbers, based upon which the operating conditions to achieve a particular charge efficiency are derived.  We subsequently present simulation results using the flow-through electrode configuration to illustrate the distribution of the local degree of intercalation, intercalation reaction rates, and salt concentration levels within RC and NRC CID cells. Using this approach we quantify the tradeoffs between energy consumption and salt removal rate with two different water sources, 500mM-NaCl and 70mM-NaCl, representative of seawater and brackish water, respectively, across a range of current densities.  In all cases we simulate cycling between −0.4 and 0.4 V cell potential limits.  We fix the dimensions of CID cells in all simulations, excluding the thickness of the diaphragm and the tank volume.  Lastly, we explore the use of diaphragms with flow configurations adjacent to electrodes, rather than through them.  

We use various metrics to quantify the desalination performance of CID devices.  Each metric is defined as a ratio of extensive quantities transferred during one cycle, comprised of one charge and one discharge step, and of the design attributes of the cycled cell.  The extensive quantities transferred include the moles of salt removed $N_{salt}$, the mass of salt removed $m_{salt}$, the volume of desalinated water produced $V_{desal}$, the net electrical energy consumed $E_{elec}$, and the moles of electrons transferred $N_{elec}$.  The metrics used include the desalination extent $\Delta c_e$, desalination energy consumption $E_d$, average salt adsorption rate (ASAR) expressed on the basis of electrode area $ASAR_a$ and IHC mass $ASAR_m$, energy normalized adsorbed salt (ENAS), and charge efficiency $\it \Gamma$.

\begin{itemize}

\item $\Delta c_e$ is the ratio of the moles of salt removed to the volume of desalinated water: $\Delta c_e = N_{salt}/V_{desal}$.

\item $E_d$ is the ratio of the net electrical energy consumed to the volume of desalinated water: $E_d=E_{elec}/V_{desal}$.

\item $ASAR_a$ is the ratio of the moles of salt removed to the product of the projected area of one electrode $A$ and the time elapsed during the cycle $t_{cyc}$: $ASAR_a=N_{salt}/(A \times t_{cyc})$.  $ASAR_m$ is the ratio of the mass of salt removed to the product of the mass of the IHC material in both electrodes $m_{IHC}$ and the time elapsed during the cycle $t_{cyc}$: $ASAR_m=m_{salt}/(m_{IHC} \times t_{cyc})$.

\item $ENAS$ is the ratio of the moles of salt removed to the net electrical energy consumed: $ENAS=N_{salt}/E_{elec}$.  Specific energy consumption is the reciprocal of ENAS.

\item $\it \Gamma$ is the ratio of the moles of salt removed to the moles of electrons transferred: $\it \Gamma=N_{salt}/N_{elec}$.  The diaphragm salt removal efficiency (DSRE) $\omega$ is equal to charge efficiency normalized by the anion transference number of the feed solution $t_-$: $\omega = \it \Gamma/t_-$.

\end{itemize}

\noindent The specific expressions used to calculate these metrics from simulated cycling data are included in the Supplementary Information.

\subsection{Effect of diaphragm design and current density}
\begin{figure}[H]
    \centering
    \includegraphics[width=\textwidth]{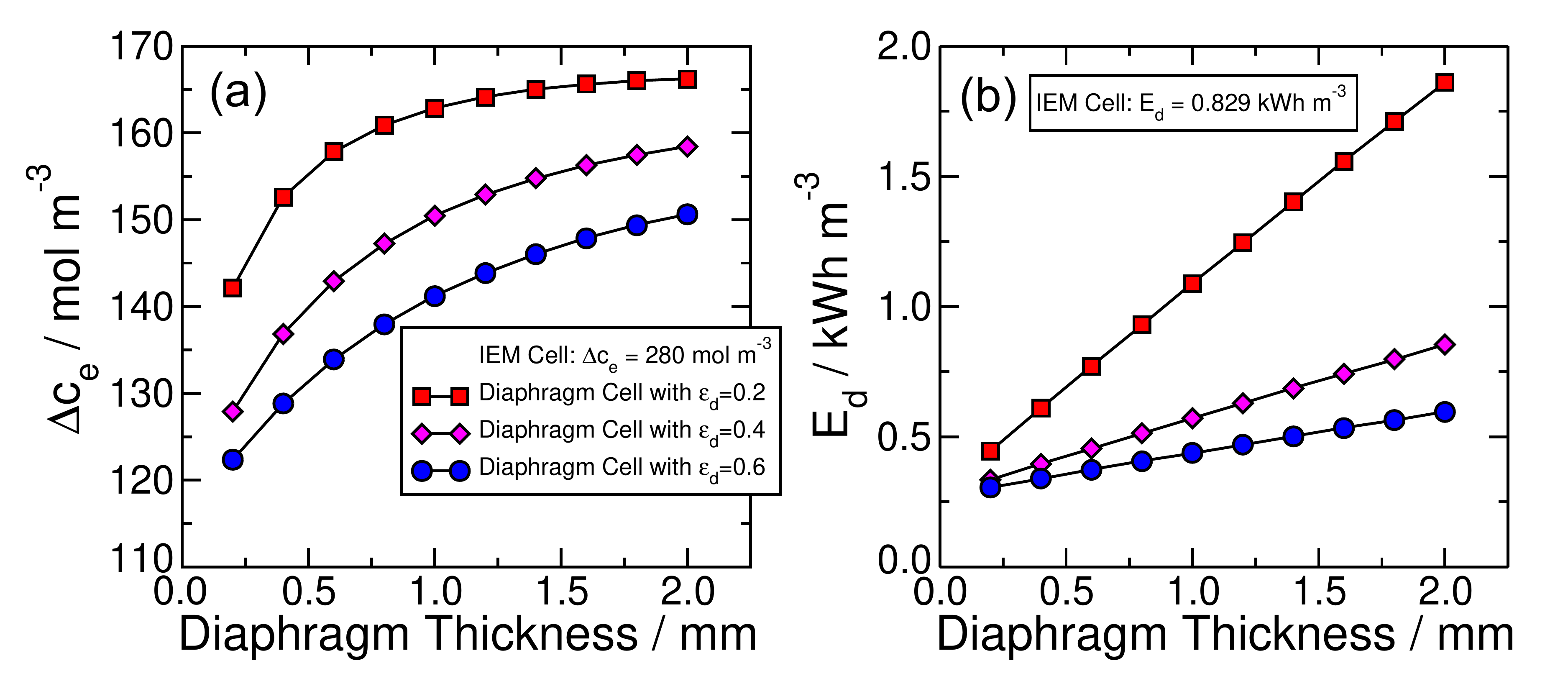}
    \caption{(a) Extent of desalination $\Delta c_e$ and (b) energy consumption per unit volume of desalinated water $E_d$ for NRC diaphragm cells as  functions of diaphragm thickness and porosity with $54\text{ A m}^{-2}$ current density.  The benchmark IEM cell case desalinates 280 $\text{mol m}^{-3}$ and consumes 0.829  $\text{kWh m}^{-3}$.}
    \label{fig:dc diaphragm-thickness}
\end{figure}

Previous simulations of CID revealed that IEMs produce greater salt removal per unit charge transferred than cells using thin, non-selective separators \cite{smith2016ion}.  Here, we predict the performance of CID cells using diaphragms of finite thickness and porosity, as shown in Fig. \ref{fig:dc diaphragm-thickness}.  In these cases 500 $\text{mol m}^{-3}$ NaCl feed solution flows through an NRC cell at $0.224$ $\text{cm min}^{-1}$ with current density fixed at $54$ $\text{A m}^{-2}$. For the present electrode thickness and NiHCF volumetric loading levels simulated, this current density corresponds to a theoretical charge time of 1 h.  For the IEM cases we model an ideally anion-selective interface across which is a Donnan potential, as in \cite{smith2017theoretical}. Under these conditions the IEM cell produces 280 $\text{mol m}^{-3}$ extent of desalination, while each of the diaphragm cells desalinate to a maximum value of 167 $\text{mol m}^{-3}$ for thick diaphragms with low porosity (Fig. \ref{fig:dc diaphragm-thickness}a).  Here, low-porosity diaphragms show increased salt removal relative to high-porosity ones because the effective chemical diffusion coefficient of salt $D_{eff}$ decreases as porosity decreases in the diaphragm, reducing the amount of ion diffusion during electrochemical cycling. However, the reduction in $\Delta c_e$ when increasing $\epsilon_d$ from 40$\%$ to 60$\%$ is marginal (10 $\text{mol m}^{-3}$), and if thicker diaphragms with higher porosity are used this marginal drop can be compensated for as a result of the increased length scale for diffusion through the diaphragm.  Considering the 280 $\text{mol m}^{-3}$ desalination extent obtained from IEM cell, these results suggest that reaching such levels of salt removal are not possible simply by changing diaphragm design \textit{ceteris paribus}. Furthermore, increased electrode thickness results in increased internal resistance, which is evidenced by increased desalination energy consumption as diaphragm thickness increases (Fig. \ref{fig:dc diaphragm-thickness}b).  This increased energy consumption also manifests in cell-potential response showing increased polarization that can reduce charge capacity utilization, charge time, and, consequently, water throughput when cycled galvanostatically between pre-specified cell-potential limits.

\begin{figure}[H]
    \centering
    \includegraphics[width=\textwidth]{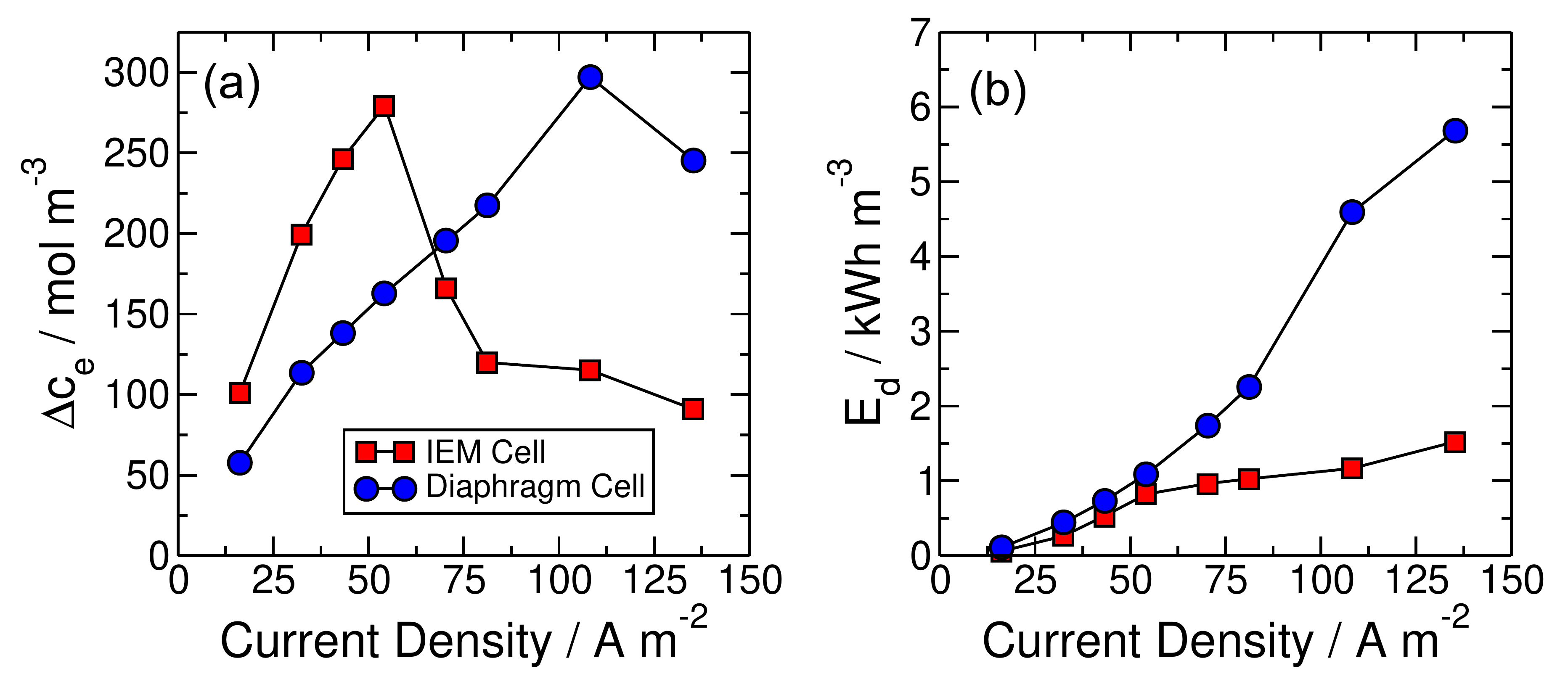}
    \caption{(a) Simulated extent of desalination and (b) energy consumption as a function of current density for the NRC diaphragm cell and a similar IEM cell.}
    \label{fig:dc-ipp aem dia}
\end{figure}

Based on the preceding findings we adopt a diaphragm design that balances salt removal and energy consumption using a certain diaphragm porosity $\epsilon_d=20\%$ and thickness $w_d=1\text{ mm}$.  Fixing these parameters, we predict the variation of desalination extent $\Delta c_e$ and energy consumption $E_d$ with current density $i$ for a fixed diaphragm design (Fig. \ref{fig:dc-ipp aem dia}).  By increasing current density above $54 \text{ A m}^{-2}$, desalination extent increases and exhibits greater salt removal than the corresponding IEM cell at $75\text{ A m}^{-2}$. For diaphragm and IEM cells transport processes in the electrolyte (due to salt depletion and ohmic polarization) limit the degree of salt removal at high current densities. In the IEM cell these transport limitations occur at lower current densities because the depletion of salt at one side of the IEM increases Donnan potential and solution resistivity. The diaphragm cell, on the other hand, transports Na$^+$ and Cl$^-$ ions continuously between electrolyte streams, resulting in less cell polarization at similar current densities.  Though the diaphragm cell produces similar maximum extent of desalination to the IEM cell (Fig. \ref{fig:dc-ipp aem dia}a), improved desalination extent comes at the cost of increased energy consumption for diaphragm cells (Fig.\ref{fig:dc-ipp aem dia}b). When both cells obtain $200\text{ mM}$ concentration drop in the effluent, the diaphragm cell and the IEM cell show specific energy consumption values of $31.7\text{ kJ mol} ^{-1}$ and $4.9\text{ kJ mol}^{-1}$, respectively.  This finding motivates the exploration of alternative strategies to simultaneously increase desalination extent while moderating energy consumption. Along these lines we subsequently explore the role of electrolyte flow rate, which affects ion transport inside the cell, and, as such, we explore the coupled roles of $i$ and $u_s$ in controlling $\Delta c_e$ in both NRC and RC diaphragm-cell architectures.

\subsection{Dimensionless performance maps of charge efficiency}

Using analytical modeling, we now assess the coupling of charge efficiency to operating conditions, device design, and material parameters.  For this model we assume that (1) a pseudo-steady salt concentration profile develops within each cell, (2) the effective chemical diffusivity of salt is a constant, (3) streamwise salt diffusion is negligible, and (4) intercalation reactions are distributed uniformly.  Under such conditions closed-form equations for effluent salt concentration can be determined by the integration of two differential equations derived from Eq.~\ref{e1} and their subsequent simplification (Appendix A).  From this analysis we extract desalination performance maps that correlate charge efficiency $\it \Gamma$ with $u_s$ and $i$, among other parameters.  We calculate a normalized version of the charge efficiency that we refer to subsequently as the diaphragm salt removal efficiency $\omega$ (DSRE).  Here the actual charge efficiency $\it \Gamma$ (defined as the moles of salt removed relative to the moles of electrons transferred) can be calculated from DSRE as $\it \Gamma=\omega t_{-}$.  For the analytical modeling results that we present in this section DSRE is equivalent to the actual desalination extent $\Delta c_{e}$ normalized by that of a cell with negligible apparent salt diffusion $\Delta c_{e,ideal}$: $\omega=\Delta c_e/\Delta c_{e,ideal}$.  Because $\omega$ is always less than unity diaphragm CID cells possess an inherent loss in charge efficiency equal to the cation transference number $t_+=1-t_-$, which for NaCl is 39$\%$ \cite{smith2016ion}, in comparison with CID cells that use anion-exchange membranes with $t_{-}=1$.  For the flow-through electrode configuration our analysis reveals that $\omega$ depends on any two of the following dimensionless parameters:

\begin{itemize}
  \item P\`{e}clet number $Pe= \frac{w_e u_s/L_{cc}}{2G_m^{\prime\prime}}$ is a characteristic ratio of the flow-driven salt transfer rate relative to the salt diffusion rate through the diaphragm, 
  \item The Damk{\"o}hler number of the first kind $Da_{I}=\frac{it_{-}/F}{w_{e}u_{s}\Delta c_{e,ideal}}$ is a characteristic ratio of the apparent salt consumption rate to the flow-driven salt transfer rate, and
  \item The Damk{\"o}hler number of the second kind $Da_{II}=Da_{I}Pe=\frac{it_{-}/F}{2\Delta c_{e,ideal}G_m^{\prime\prime}}$ is a characteristic ratio of the \text{apparent} salt consumption due to intercalation reactions relative to the salt diffusion rate through the diaphragm.
\end{itemize}

\noindent For flow-by and flow-behind configurations these dimensionless parameters can be determined by substituting the desalinating volumetric flow rate per unit cell depth $L_d$ for $w_eu_s$ in the expressions above. In these expressions $G_m^{\prime\prime}$ is salt conductance per unit diaphragm area (in units of  $\text{m s}^{-1}$) that produces a molar salt flux at any given streamwise location equal to $N^{\prime\prime}=G_m^{\prime\prime}(c_e^+-c_e^-)$, where $c_e^{+/-}$ is the salt concentration in the positive/negative electrode at the corresponding streamwise position.  Considering the mass transfer resistance due to the diaphragm only, $G_m^{\prime\prime}=D_{eff,d}/w_d$; the additional mass transfer resistance of the porous electrodes themselves can also be included, as described in Appendix A.  Using our analytical model $\omega$ can be expressed in terms of $Pe$ and $Da_{II}$:

\begin{equation}
    \omega=Pe\big(1-\text{exp}(-1/Pe)\big) \text{ for NRC and}
    \label{e6}
\end{equation}
\begin{equation}
    \omega=Da_{II}\Big(1-\text{exp}\Big(\frac{Pe}{Da_{II}}\big(\text{exp}(-1/Pe)-1\big)\Big)\Big) \text{}
    \label{e7}
\end{equation}

\noindent  for RC cells. Here, the expression for RC cells neglects the volume of feedwater stored within electrodes and is therefore strictly valid in the limit of infinite tank size.  

\begin{figure}[H]
    \centering
    \includegraphics[width=\textwidth]{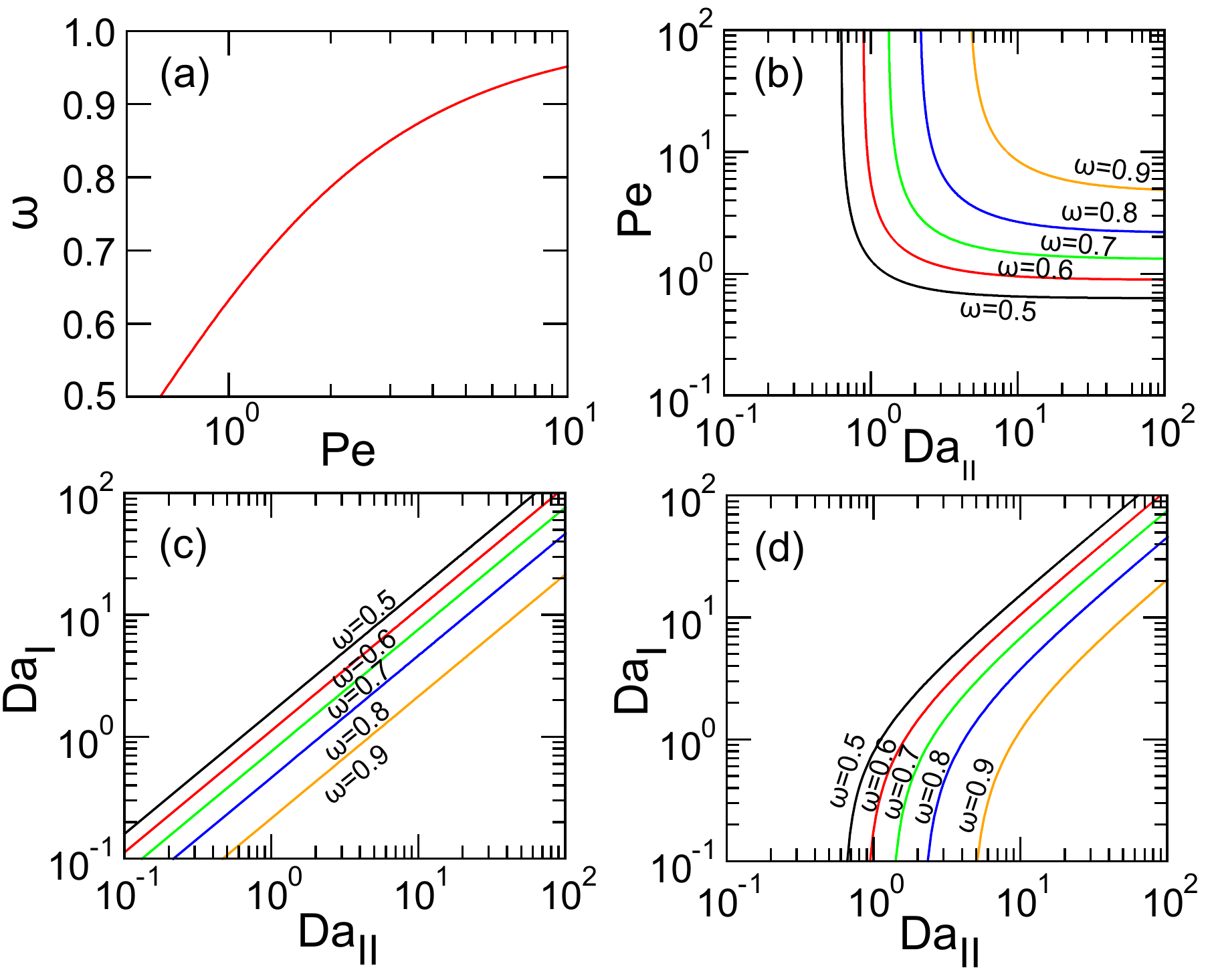}
    \caption{(a) DSRE as a function of $\it Pe$ for the non-recycling cell and (b) contours of DSRE in the space of $\it Pe$ versus $\it Da_{II}$ for the recycling cell. Contours of DSRE in the space of $\it Da_{I}$ versus $\it Da_{II}$ (c) for the non-recycling diaphragm cell and (d) for the recycling cell.}
    \label{fig:Pe-Daii}
\end{figure}

We now examine the limits of $\omega$ that are achievable as a function of $Pe$, $Da_I$, and $Da_{II}$.  Figure \ref{fig:Pe-Daii}a shows $\omega$ for the NRC cell, which depends on $Pe$ alone and not explicitly on $Da_{II}$.  This effect occurs because the DSRE for the NRC cell depends strictly on the rate of salt advection, relative to the rate of salt diffusion across the diaphragm.  Hence, above a threshold value of $Pe=4$, 90$\%$ DSRE is obtainable.  In contrast, when an RC cell is used, the effect of intercalation reaction rates (i.e., current density), diffusion across the diaphragm, and salt advection become decoupled.  This is evidenced in Fig.~\ref{fig:Pe-Daii}b where DSRE depends on both $Pe$ and $Da_{II}$.  For slow reaction rates, corresponding to small $Da_{II}$ values, $\omega$ depends only on $Da_{II}$, while for fast reaction rates (large $Da_{II}$) $\omega$ depends only on $Pe$.  The former condition arises in the limit that the rate of advection is so fast that salt concentration is practically uniform within a given electrode at any given instant in time.  As such, DSRE is limited by the time required to finish desalinating, which is inversely proportional to $Da_{II}$.  In the opposite limit (high $Da_{II}$) advective transport limits DSRE, as in the case of the NRC cell irrespective of the magnitude of current density or, equivalently, $Da_{II}$.

We also examine DSRE in the space of $Da_I$ and $Da_{II}$ (Figs.~\ref{fig:Pe-Daii}c-d).  Such maps are meaningful to explore the tradeoffs between diaphragm design and flow speed when current density is fixed: increasing diaphragm effectiveness by suppressing $G^{\prime\prime}_m$ is captured by decreases in $Da_{II}$, while increasing flow speed is captured by decreasing $Da_I$.  For the NRC cell the contours of $\omega$ show that, when diaphragm conductance decreases ($Da_{II}$ increases) flow speed can decrease in proportion ($Da_I$ increases) and the same level of DSRE is attainable.  In contrast, for the RC cell the effect of flow speed (or $Da_I$) diminishes for high flow velocities (small $Da_I$), the achievable DSRE level depends entirely on the ratio of intercalation current to diaphragm salt conductance.

\subsection{Simulated performance for different water sources and cell architectures}

Ultimately, the efficient operation of a diaphragm-based desalination device will depend not only on the salt-removal efficiency but also on the energy consumed during desalination.  Energy consumption results from the sources of polarization arising from local transport processes within porous electrodes and within diaphragms, and, as such, we use a two-dimensional, transient porous electrode model to capture such effects.  With this model we solve for the distribution of intercalation reaction rates within each electrode subject to electrochemical kinetics, rather than assuming it to be uniform \textit{a priori} as in the analytical model.  Furthermore, the analytical modeling results presented in the preceding section show that charge efficiency depends on cell architecture, operating conditions, and on the desired extent of desalination.  Accordingly, we present simulated results with influent salinity levels representative of brackish ($c_e^0=$ 70 mM) and seawater ($c_e^0=$ 500mM) using both NRC and RC cell architectures.

When choosing what operating conditions to use in the simulation of various influent sources and cell architectures, we consider what operating conditions are likely to minimize energy consumption for the RC cell architecture.  Using Fig.~\ref{fig:Pe-Daii}b we showed that, in the limit of high flow speed, DSRE depends only on the applied current density when diaphragm design is fixed. In this limit applied current density is minimized along a particular DSRE contour, and, hence, ohmic contributions to polarization are expected to be smallest.  Thus, in all subsequent simulations we choose certain $\omega$ and $\Delta c_{e,ideal}$ values, and we determine what current density must be applied to achieve these levels in the high flow-rate limit with an RC cell using the analytical model.  For these cases we choose $\Delta c_{e,ideal}$ to produce 60$\%$ degree of desalination (i.e., $\Delta c_{e,ideal}=0.6c_e^0$).  We use the same conditions for the NRC cell, and we choose the tank volume for the RC cell equal to the desalinated effluent volume for the NRC at the same current density $i$, the same DSRE $\omega$, and cycled with the theoretical capacity of NiHCF. For the corresponding simulations that use these operating conditions, we fix the tank volume for the RC cell to $5.87$ and $0.821$ cm$^3$ per cm of cell depth ``into the page'' in brackish water cases and seawater cases, respectively.  In all cases we assumed a salt conductance of $1.07\times10^{-7} \text{m s}^{-1}$ based on the dilute solution bulk salt diffusivity of $1.61\times10^{-9} \text{m}^2 \text{ s}^{-1}$ \cite{rard1979mutual} and the $G_m^{\prime\prime}$ expression in Appendix A that includes the effects of mass transfer resistance from both the diaphragm and the electrodes.

Table \ref{t1} shows the results of simulations at various current density $i$ and DSRE values set using the analytical model $\omega_{set}$ and that resulted in simulated DSRE values $\omega_{sim}$.  Inspection of these data reveal that faster flow speeds are required to run RC cells at a certain DSRE level than for the corresponding NRC cells.  RC cells require high flow speeds to overcome salt diffusion through diaphragms.  In contrast, for the NRC cell low flow speeds are used, corresponding to long residence times which; produce lower polarization and longer charge times; as we show later.  Table \ref{t1} also shows $\omega_{sim}$, $\Delta c_e$, and $E_d$ for the simulations.  Here, we emphasize that $\omega_{sim}$ was calculated as charge efficiency divided by anion transference number, $\it \Gamma/t_{-}$.  The simulated DSRE values for the NRC cell were 7$\%$ and 4$\%$ lower than the DSRE values predicted with the analytical model $\omega_{set}$ for seawater and brackish water simulations, respectively.  For the RC cell the discrepancy between simulation and the model is 9.3$\%$ on average for seawater and 2$\%$ for brackish water. This discrepancy is caused, in part, by large polarization experienced by the RC cell at high $i$ and $u_s$. Brackish water DSRE agrees better with analytical model values because $i$ and $u_s$ are smaller than in seawater simulations. As we show later, such conditions promote more uniform distribution of salt within electrodes, which approaches the idealized concentration distribution assumed in the analytical model.  In all cases, the extent of desalination is influenced by the simulated DSRE, where NRC cells produce greater desalination extent than RC cells with the same current density.  The desalination energy consumption also differs between cell architectures, in part due to the different desalination extent achieved and due to the electrochemical kinetics and transport processes occurring within the electrodes. In terms of specific energy consumption, the average energy consumption of brackish water simulation ($14.4$ kJ mol$^{-1}$) is higher than the seawater simulation ($13.1$ kJ mol$^{-1}$). When the two kinds of cells are operating with similar $\omega_{sim}$, the RC cell generally needs less energy than NRC cell. For instance, RC cell and NRC cell consume $34$ kJ mol$^{-1}$ and $36$ kJ mol$^{-1}$ separately when they are operated with seawater and a DSRE level of 0.83.

\begin{table}[H]
\centering
\caption{Results for the simulation of brackish and seawater desalination using both NRC and RC diaphragm cell architectures.}
\label{t1}
\normalsize
\setlength{\arrayrulewidth}{.18em}
\resizebox{\textwidth}{!}{
{\renewcommand{\arraystretch}{1}
\begin{tabular}{cccccccccccc}
\hline
i/A m$^{-2}$ & $\omega_{set}$ & $u_s$/cm min$^{-1}$ & $\omega_{sim}$ & $\Delta c_e$/mM & $E_d$/kWh m$^{-3}$ &                      & $\omega_{set}$ & $u_s$/cm min$^{-1}$ & $\omega_{sim}$ & $\Delta c_e$/mM & $E_d$/kWh m$^{-3}$ \\ \hline
\multicolumn{6}{c}{\textbf{NRC Cell with Brackish Water}}                                                   & \multicolumn{1}{l}{} & \multicolumn{5}{c}{\textbf{RC Cell with Brackish Water}}                                     \\
17.0         & 0.950          & 0.307               & 0.918          & 38.90           & 0.529              &                      & 0.950          & 5.08                & 0.948          & 31.1            & 0.372              \\
10.0         & 0.920          & 0.181               & 0.887          & 37.45           & 0.328              &                      & 0.920          & 9.63                & 0.902          & 36.8            & 0.309              \\
8.0          & 0.900          & 0.145               & 0.866          & 36.54           & 0.268              &                      & 0.900          & 4.11                & 0.883          & 36.7            & 0.263              \\
6.0          & 0.870          & 0.108               & 0.831          & 35.04           & 0.209              &                      & 0.870          & 3.29                & 0.853          & 35.8            & 0.207              \\
5.0          & 0.850          & 0.090               & 0.803          & 33.88           & 0.178              &                      & 0.850          & 6.79                & 0.831          & 35.0            & 0.177              \\
\multicolumn{6}{c}{\textbf{NRC Cell with Seawater}}                                                         &                      & \multicolumn{5}{c}{\textbf{RC Cell with Seawater}}                                           \\
75.0         & 0.920          & 0.190               & 0.829          & 258             & 2.57               &                      & 0.920          & 123                 & 0.832          & 246             & 2.31               \\
57.0         & 0.900          & 0.144               & 0.814          & 253             & 2.01               &                      & 0.900          & 10.9                & 0.796          & 244             & 1.87               \\
49.0         & 0.880          & 0.126               & 0.822          & 250             & 1.76               &                      & 0.880          & 5.66                & 0.795          & 241             & 1.66               \\
40.0         & 0.840          & 0.101               & 0.786          & 243             & 1.49               &                      & 0.840          & 3.83                & 0.762          & 236             & 1.40               \\
35.0         & 0.820          & 0.089               & 0.773          & 239             & 1.33               &                      & 0.820          & 6.79                & 0.750          & 232             & 1.24               \\ \hline
\end{tabular}}}
\end{table}

\begin{figure}[H]
    \centering
    \includegraphics[width=\textwidth]{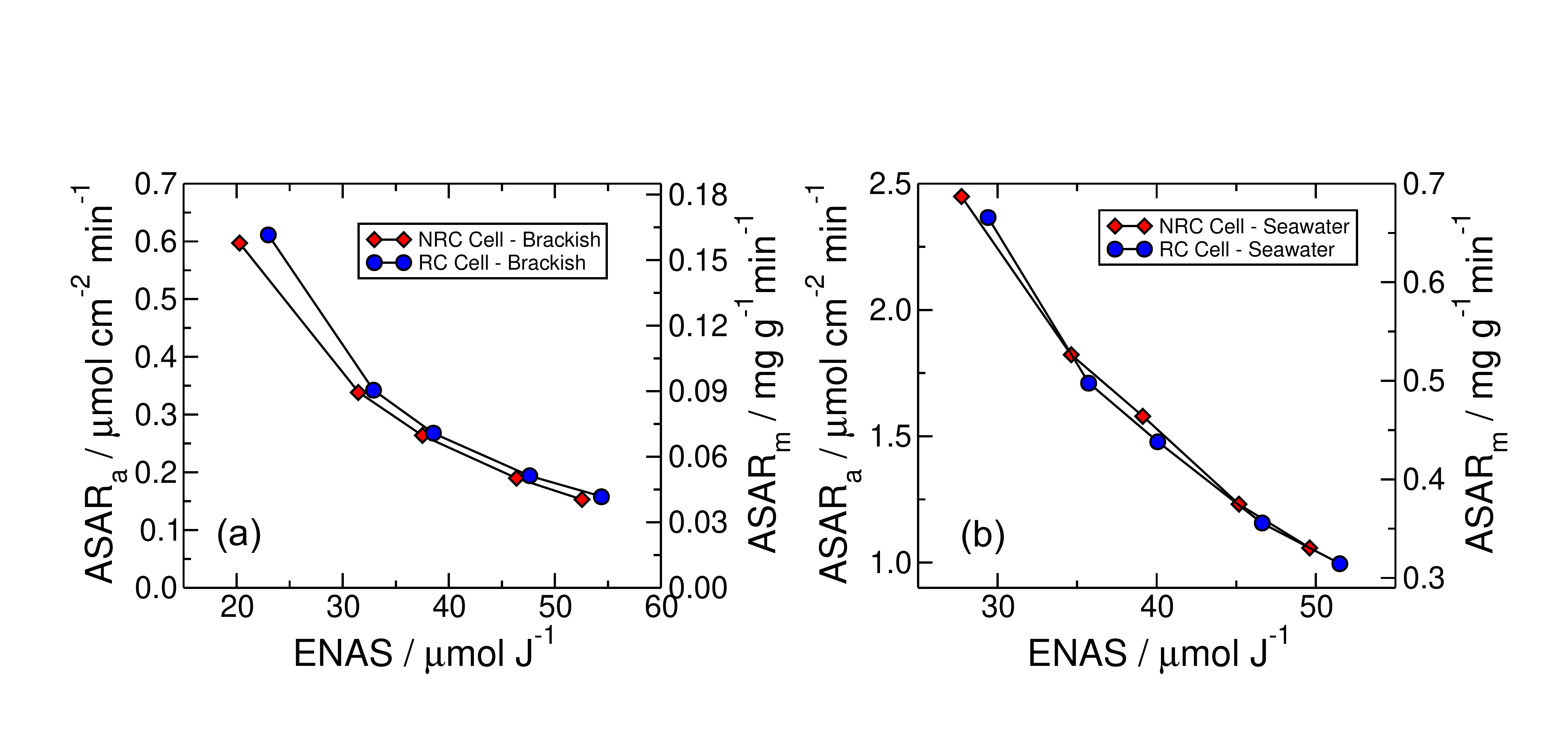}
    \caption{ASAR versus ENAS for (a) brackish salinity and (b) seawater salinity cases from the results in Table~\ref{t1}.}
    \label{fig:ASAR-ENAS}
\end{figure}

We now quantify the tradeoffs between ASAR and ENAS for the cases described in Tables~\ref{t1}. ASAR is a measure of the desalination rate and ENAS is a measure of the salt removed per unit electrical energy input.  In general, a process which has both high ASAR and ENAS (i.e., the top right corner of the Figs.~\ref{fig:ASAR-ENAS}a,b) is preferred. However, Fig.~\ref{fig:ASAR-ENAS} shows that, irrespective of feedwater salinity and cell architecture (i.e., use of either the RC or NRC cell), an energy efficient process (high ENAS) is only achieved at slow desalination rates (low ASAR).  On the other hand, high desalination rates (high ASAR) have low energy efficiency (low ENAS). That said, the RC cell, for most cases simulated here, is more energy efficient than the NRC cell when both cells have similar ASAR. The exception is found when both cells desalinate seawater with high energy efficiency (high ENAS). Meanwhile, when a fixed value of ENAS is obtained from these cells, the RC cell tends to desalinate influent faster than NRC cell. We also note the influence of different source water on cell performance, where the magnitude of ASAR is larger for seawater simulation cases at the same ENAS as brackish water.

\begin{figure}[H]
    \centering
    \includegraphics[width=\textwidth]{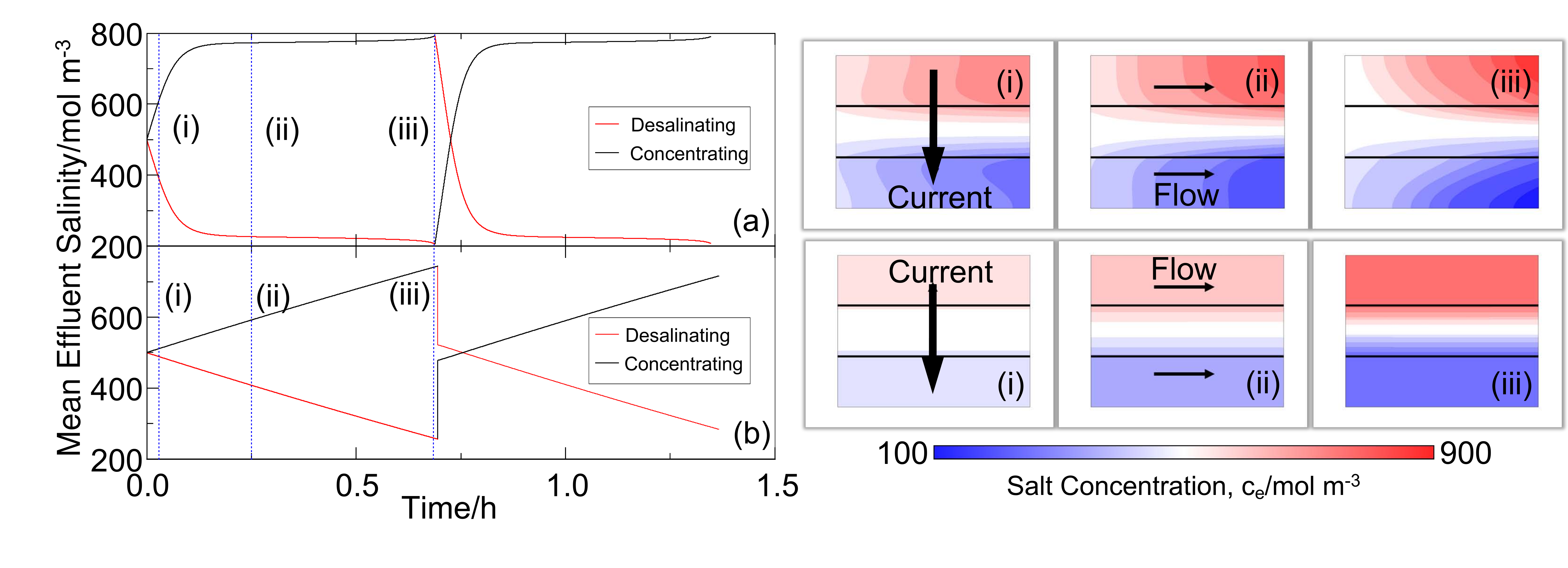}
    \caption{Space-averaged effluent salt-concentration (left) during the charge/discharge cycle of (a) an NRC cell with $i=75$ $\text{A m}^{-2}$ and $u_s=0.190$ $\text{cm min}^{-1}$ and (b) an RC cell with $i=75$ $\text{A m}^{-2}$ and $u_s=123$ $\text{cm min}^{-1}$ operating with seawater-level influent salinity. At the indicated instants in time (i-iii) the distribution of NaCl in the electrolyte is shown (right).}
    \label{fig:mean effluent seawater}
\end{figure}

\begin{figure}[H]
    \centering
    \includegraphics[width=\textwidth]{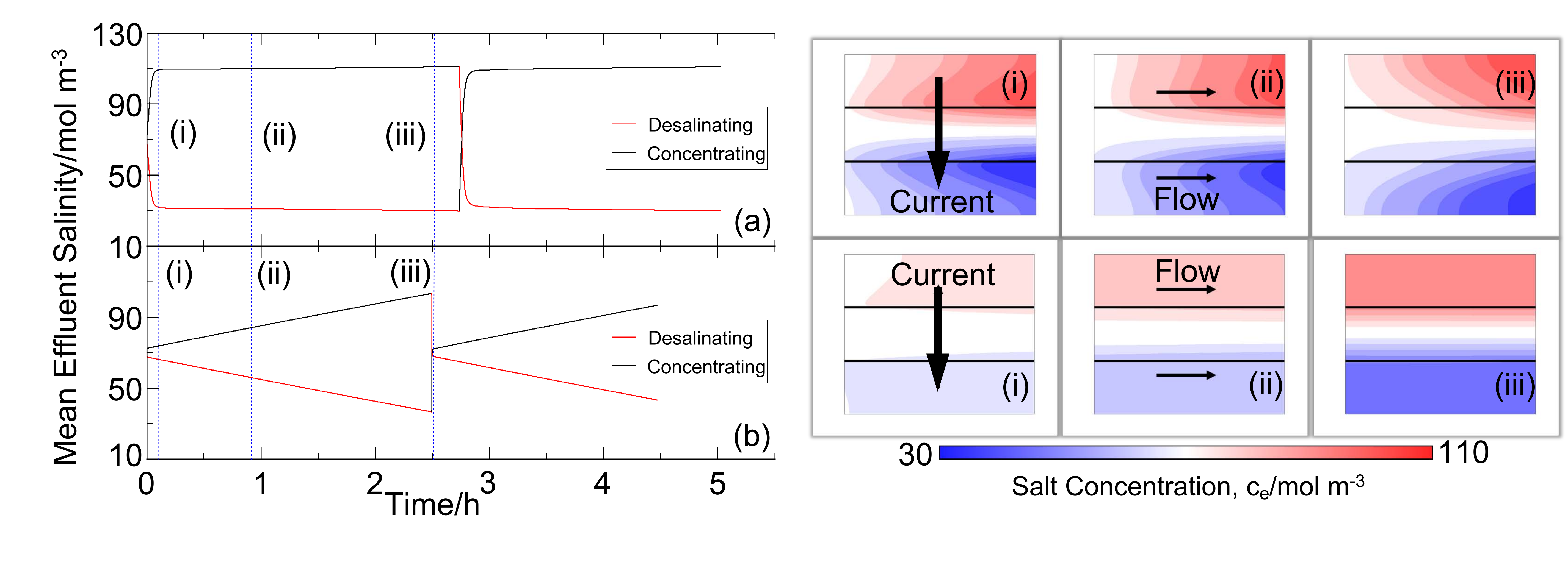}
    \caption{Space-averaged effluent salt-concentration (left) during the charge/discharge cycle of (a) an NRC cell with $i=17$ $\text{A m}^{-2}$ and $u_s=0.307$ $\text{cm min}^{-1}$ and (b) an RC cell with $i=17$ $\text{A m}^{-2}$ and $u_s=5.08$ $\text{cm min}^{-1}$ operating with brackish-level influent salinity. At the indicated instants in time (i-iii) the distribution of NaCl in the electrolyte is shown (right).}
    \label{fig:mean effluent brackish water}
\end{figure}

In what follows we elucidate the transport and kinetic mechanisms that produce the observed DSREs in our simulations.  We first examine the variations of salt concentration within diaphragm-based CID cells during charge and discharge.  Figures \ref{fig:mean effluent seawater} and \ref{fig:mean effluent brackish water} show two-dimensional maps of salt concentration at different instants in time for operation with sea and brackish water level NaCl concentrations and effluent salt concentration as a function of time using both NRC and RC cell architectures.  The effluent salinity of NRC cells in both figures reaches a pseudo-steady state within the first fifth of its charging time, while the RC cell, on the other hand, never reaches steady state and instead shows continuously increasing effluent salinity until the terminal cutoff potential is reached. In comparison with the RC cell, the NRC cell exhibits streamwise salt concentration variations that are much more significant because desalination/concentration is performed in a single pass using the NRC cell architecture.  The RC cell also shows larger desalinated effluent concentration than the NRC cell at the end of both charge and discharge steps because it charges/discharges for a shorter period of time than the NRC cell, an effect that we later attribute to cell polarization during galvanostatic cycling.

Examination of the $c_e$ distributions at time instants (i), (ii), and (iii) reveals that, for the NRC cell significant variations in salt concentration persist in both the streamwise and thru-plane directions.  In contrast, RC cells operated at the same current density as NRC cells show salt concentration distributions that are practically one-dimensional, varying primarily in the thru-plane direction along which electric current is applied to the cell.  As we show subsequently, these effects influence the distribution of intercalation reaction rates within each of the porous electrodes, the transport of salt and current through the diaphragm, and the energy consumed during desalination.


Our previous theoretical investigations with IEM-based CID cells \cite{smith2016ion,smith2017theoretical} revealed that intercalation reactions propagate in the streamwise direction when a flow-configuration is used, and, as a result, intercalation reaction rates are not uniform.  Here, we show that stream-wise salt concentration gradients are directly responsible for those effects as well as the distribution of intercalation reaction rates.  We demonstrate this by considering the expression for current density $\vec{i_e}$ in the present binary electrolyte \cite{smith2017theoretical}: $\vec{i_e}=-\kappa_{eff}(\nabla{\phi_e}-\frac{2R_gT}{F}t_-\gamma_{\pm}\nabla{\ln{c_e}})$, where $\phi_e$ is the solution-phase potential (defined as the reduced electrochemical potential of cations in solution \cite{lai2011mathematical}) and $\gamma_{\pm}$ is the thermodynamic factor accounting for deviations from ideal activity in solution \cite{lai2011mathematical}.  When electrolyte current density, kinetic overpotential, solid-phase potential differences, and  gradients of $x_{Na}$ within IHC particles are all negligible we can derive an expression (see Appendix B) relating the intercalated-Na fraction at points 1 and 2 in a porous electrode ($x_{Na,1}$ and $x_{Na,2}$) to the salt concentration in the electrolyte at the same points ($c_{e,1}$ and $c_{e,2}$): 

\begin{equation}
    c_{e,2}/c_{e,1}=\bigg\{\frac{(1-x_{Na,1})x_{Na,2}}{(1-x_{Na,2})x_{Na,1}}\bigg\}^{1/(2t_-)}.
\end{equation}

\begin{figure}[H]
    \centering
    \includegraphics[width=\textwidth]{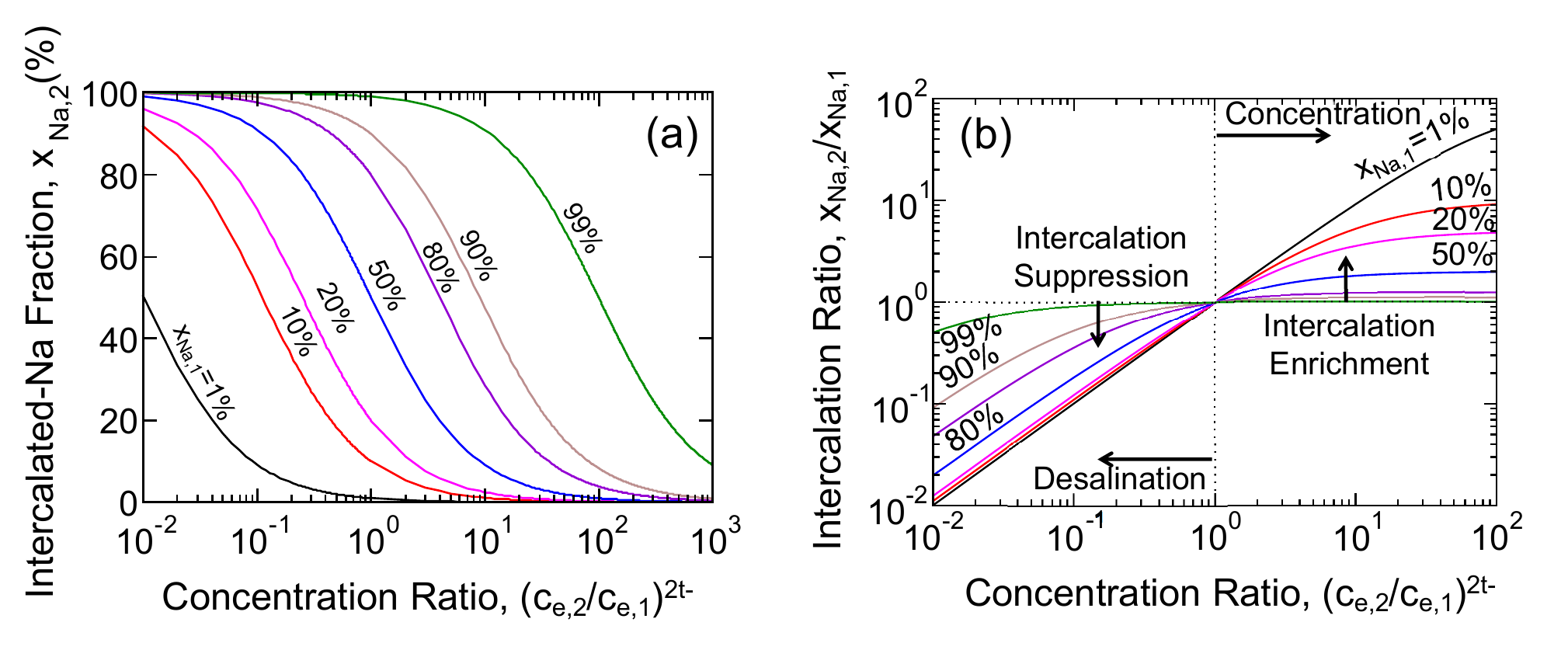}
    \caption{Theoretical predictions of the intercalated-Na fractions at two points in a porous electrode for certain salt concentrations.}
    \label{fig:diff_pot_eq}
\end{figure}

\noindent The above expression assumes ideal solution behavior in the electrolyte ($\gamma_{\pm}=1$, unlike in the present simulations where we consider concentrated solution $\gamma_{\pm}$ values) and regular solution behavior in IHC particles (as assumed in the present simulations and in Ref.~\cite{smith2017theoretical}).  The resulting salt concentration ratio $c_{e,2}/c_{e,1}$ that produces a certain difference in intercalated-Na fraction between the same points is shown in Fig.~\ref{fig:diff_pot_eq}a, b.  For the sake of illustration consider a scenario relevant to the negative electrode of an NRC cell during charging, where salt concentration is low at the electrode's outlet and it is high at the electrode's inlet because that electrode desalinates.  Figure \ref{fig:diff_pot_eq} shows that if IHC material near the inlet is Na-deficient (e.g., $x_{Na,1}=1\%$) IHC material near the inlet must have even greater degree-of-intercalation since the negative electrode would desalinate salt in such a scenario (i.e., because $c_{e,2}/c_{e,1}<1$). Hence, it is expected that intercalation reactions would be favored near the inlet of the negative electrode in such a scenario.  

\begin{figure}[H]
    \centering
    \includegraphics[width=\textwidth]{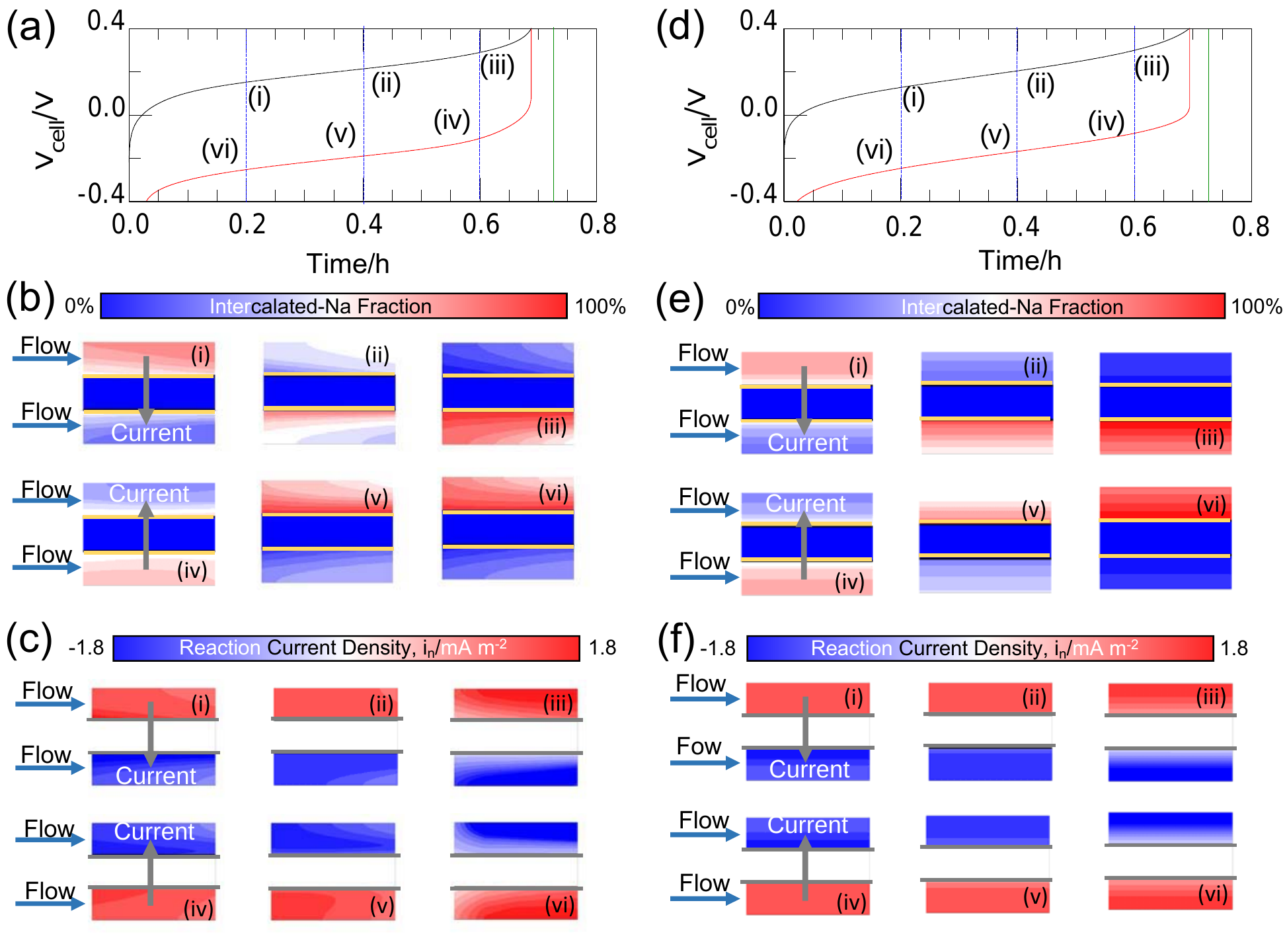}
    \caption{Cell potential as a function of time for the first charge/discharge cycle of (a) an NRC cell with $i=75$ $\text{A m}^{-2}$ and $u_s=0.190$ $\text{cm min}^{-1}$, and (d) an RC cell with $i=75$ $\text{A m}^{-2}$ and $u_s=123$ $\text{cm min}^{-1}$ with seawater-level influent salinity.  At the indicated instants in time (i-vi), snapshots of intercalated-Na fraction inside (b) the NRC cell, and (e) RC cell are shown, in addition to those of the reaction current density $i_n$ for (c) the NRC cell and (f) the RC. The solid green lines in (a) and (d) indicate the theoretical charging time and the yellow and grey horizontal lines in the distribution plots show the diaphragm edges.}
    \label{fig:seawater snapshot}
\end{figure}

\begin{figure}[H]
    \centering
    \includegraphics[width=\textwidth]{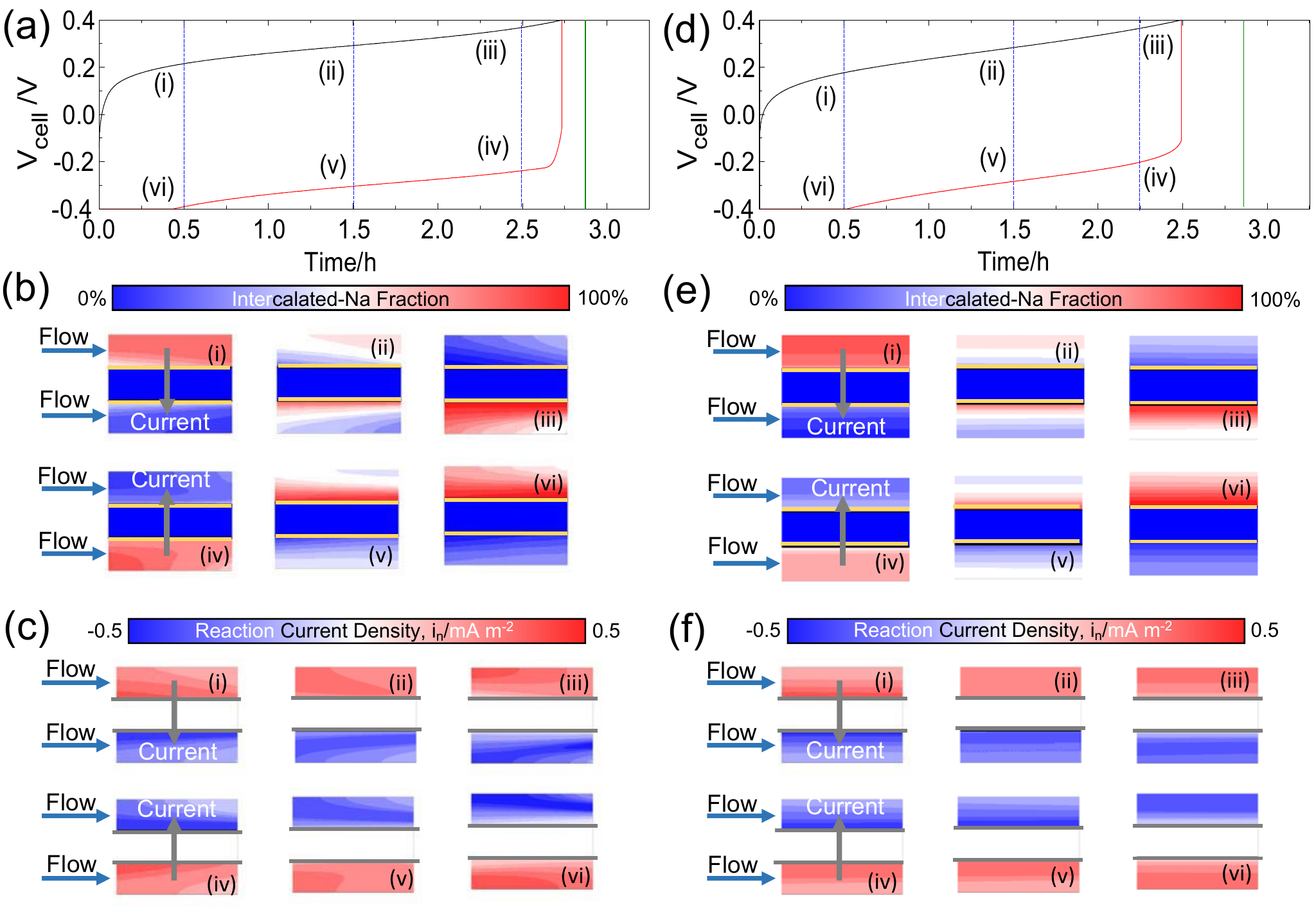}
    \caption{Cell potential as a function of time for the first charge/discharge cycle of (a) an NRC cell with $i=17$ $\text{A m}m^{-2}$ and $u_s=0.307$ $\text{cm min}^{-1}$, and (d) an RC cell with $i=17$ $A\text{ }m^{-2}$ and $u_s=5.08$ $\text{cm min}^{-1}$ with brackish-level influent salinity.  At the indicated instants in time (i-vi), snapshots of intercalated-Na fraction inside (b) the NRC cell, and (e) RC cell are shown, in addition to those of the reaction current density $i_n$ for (c) the NRC cell and (f) the RC. The solid green lines in (a) and (d) indicate the theoretical charging time and the yellow horizontal lines in the distribution plots show the diaphragm edges.}    
    \label{fig:brackish snapshot}
\end{figure}

We now examine the distributions of intercalated-Na fraction and intercalation reaction rates in space during charge and discharge of diaphragm-based CID cells, as well as the corresponding variation of cell potential with time, for seawater-level salt concentrations in Fig.~\ref{fig:seawater snapshot} and for brackish-level salt concentrations in Fig.~\ref{fig:brackish snapshot}. We note that, though continuous distributions of $x_{Na}$ are presented in these figures, the electrodes are microscopically heterogeneous.  Hence, $x_{Na}$ at any given point predicted with our homogenized approach is representative of NiHCF particles surrounded by electrolyte and electron conductor at that point. When both NRC and RC cells are operated with seawater-level influent salinity (Figs.~\ref{fig:seawater snapshot}a and \ref{fig:seawater snapshot}d, respectively), both architectures show polarization of approximately 200 mV, whereas when brackish-level influent salinity is used the RC cell (Fig.~\ref{fig:brackish snapshot}d) shows substantially greater polarization than the NRC cell (Fig.~\ref{fig:brackish snapshot}a) and consequently shorter charge/discharge times during galvanostatic cycling.  Because polarization reduces the achievable charge/discharge time during galvanostatic cycling between two potential limits, this polarization limits the achievable extent of desalination in RC cell configurations, where the effluent salinity continuously increases with time.  We also note that there are differences in the polarization experienced by cells with brackish versus seawater NaCl levels (i.e., when comparing Figs.\ref{fig:brackish snapshot}a,d with Figs.\ref{fig:seawater snapshot}a,d).  Brackish water simulations possess lower influent salt concentration that increases internal resistance and produces more significant polarization.  For brackish simulations the RC cell produces larger polarization than the NRC cell, and, as a result, it cycles galvanostatically between two potential limits with shorter charge/discharge times.

Our simulated results also confirm the theoretical finding that non-uniformity in the distribution of intercalation reactions is correlated with the uniformity of salt concentration within an electrode at any given time.  For the present NRC cells during the charging process of the negative electrode, wherein IHC intercalates and salt dilutes, the ratio of $c_e$ at the outlet relative to the inlet $c_{e,out}/c_{e,in}$ is 0.40 for a degree-of-desalination equal to 60$\%$.  Since IHC particles in the negative electrode begin the charging step in a Na-deficient state (i.e., $x_{Na}\sim0$) IHC particles near the outlet intercalate Na$^+$ preferentially because $x_{Na,out}/x_{Na,in}$ is less than unity when $c_{e,out}/c_{e,in}$ is less than unity (see Fig.~\ref{fig:diff_pot_eq}b).   In the extreme limit where $x_{Na,in}\rightarrow0$ we find that $x_{Na,out}/x_{Na,in}$ approaches $(c_{e,out}/c_{e,in})^{2t_-}$.  For the present NaCl electrolyte with $c_{e,out}/c_{e,in}=0.4$ in the desalinating electrode, this theory suggests threefold faster intercalation near the inlet than at the outlet of an NRC cell architecture.  Our simulated results in Figs.~\ref{fig:seawater snapshot}b,c and e,f reveal that intercalation reaction rates are focused near the inlet of the desalinating electrode, confirming this qualitative theoretical finding. 

For the positive electrode during discharge, wherein IHC de-intercalates and salt concentrates, $c_{e,out}/c_{e,in}$ is greater than unity and IHC material starts in a Na-rich state (i.e., $x_{Na}\sim1$).  Our equilibrium theory predicts that in the Na-rich limit $x_{Na,2}/x_{Na,1}$ is practically unity for all salt concentrating conditions (see Fig.~\ref{fig:diff_pot_eq}b), suggesting that there is no preference for the localization of de-intercalation processes  within the concentrating electrode.  This finding contrasts with our observations of simulated reaction distributions in the concentrating electrodes of NRC cells.  Those simulated results show higher rates of de-intercalation near the inlet of each electrode (Figs.~\ref{fig:seawater snapshot}b,c and ~\ref{fig:brackish snapshot}b,c).  Because our simulations account for the influence of finite-rate transport effects, in addition to equilibrium potential, we attribute their simulated localization in the concentrating electrode to the ohmic transport limitation induced by localization of reactions within the desalinating electrode.  When the NRC cell begins to discharge, the same reaction variations exist but are interchanged between electrodes.

For RC cells with brackish and seawater influent NaCl levels streamwise variations in reaction rates are eliminated altogether as a result of the small difference in salt concentration between the inlet and outlet of the electrodes (see Figs.~\ref{fig:seawater snapshot}e,f and ~\ref{fig:brackish snapshot}e,f), consistent with expectations from our diffusion-potential equilibrium theory.  Thus, the RC cell configuration mitigates reaction non-uniformities, and this effect could help to prevent capacity-fade processes that could occur in practical cell operation, though not modeled here.  Despite this finding, there are still thru-plane (X-direction) gradients in the degree of intercalation that develop as a result of the finite electrolyte-phase potential-difference induced as a result of the finite electrolyte current density transferred through the porous electrodes into the diaphragm.

\subsection{Effect of Flow Configuration}

The aforementioned results were obtained using a diaphragm CID cell with flow-through (FT) electrodes only.  While such a flow configuration has found use in redox flow batteries \cite{darling2014influence} and its implementation is the subject of our own on-going research, it has yet to be demonstrated experimentally for CID cells.  In our previous work \cite{smith2017theoretical}, different flow configurations were shown to affect CID device polarization, desalination rate, and energy consumption. In IEM-based CID cells, the flow-through configuration (FT) produces lowest cell polarization as a result of the absence of open channels, and recovers more energy during the discharging cycle \cite{smith2017theoretical}. Here, we also consider two other flow configurations: a flow-by (FB) design in which the influent is pumped through open channels between electrodes and the diaphragm (as in Ref.~\cite{smith2017theoretical}) and a flow-behind (FBH) design similar to the former, except that the open channels are located between current collectors and electrodes (as in Ref.~\cite{ Porada2017}).  Since an exhaustive numerical investigation of flow configurations with diaphragm-based CID could constitute a complete study in its own right, here we compare the cycling dynamics amongst these three flow configurations for a single case of cell operation using brackish water, $17 \text{A m}^{-2}$ current density, and $1.84$ cm$^3$ h$^{-1}$ volumetric flow rate per cm of cell depth using an NRC architecture.

\begin{figure}[H]
    \centering
    \includegraphics[width=\textwidth]{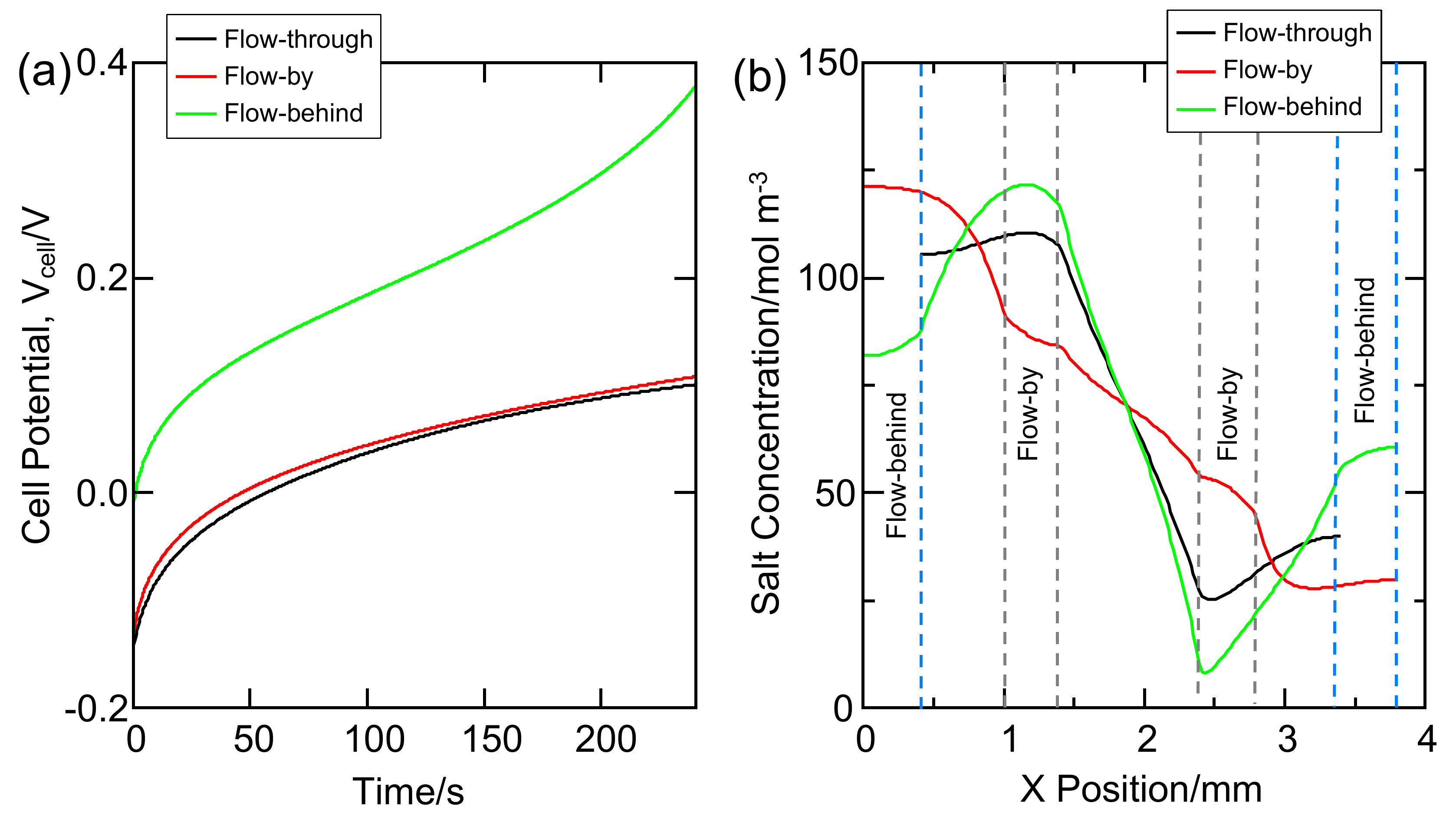}
    \caption{(a) Cell potential variation versus time and (b) salt concentration distribution along the $X$ direction at the outlet of the cell for flow-through, flow-behind, and flow-by configurations. The blue and gray dashed lines in (b) indicate the locations of flow channels for the flow-behind and the flow-by configuration, respectively.}    
    \label{fig:flow-behind results}
\end{figure}

Figure \ref{fig:flow-behind results}(a) shows the variation of cell potential with time for FBH, FT, and FB configurations for the first six minutes of cell charging.  While FT and FB configurations exhibit similar cell potential curves, the FBH configuration exhibits substantially more polarization that results in it reaching the 0.4 V cutoff very early in the charging process (6 min compared with a theoretical charge time of 4 h).  To elucidate why this effect occurs we plotted the concentration profile in the $X$ (thru-plane) direction at the outlet of each of the three cells (Fig. \ref{fig:flow-behind results}(b)).  FT obtains the most uniform concentration distribution within the electrodes.  Although intercalation reactions occur inside of the electrodes in all three flow configurations, salt diffusion limits the rate at which current can flow through the cell for FB and FBH configurations.  This effect is most clearly evident for the FBH configuration that exhibits local depletion of salt to 8 mM at the electrode edge nearest the diaphragm.  The concentration variations within the electrodes of FB and FBH cells reduce the extent to which the feed water stream can be desalinated.  As a result, the FT cell produces higher desalination extent than the other arrangements.  Considering all three flow configurations with the same electrode thickness and operating conditions, the performance of the FT configuration serves as an upper bound for FB and FBH.  Despite the finding here that FB and FBH configurations exhibit poorer performance than FT, the electrode thickness and operating conditions for these configurations can be further optimized to improve performance when a diaphragm is used to separate the electrodes.

\section{Conclusions}

A membrane-free cation intercalation desalination device using a porous diaphragm as a separator is simulated here to quantify the tradeoffs between energy consumption and salt removal rate. The use of a diaphragm is primarily motivated by its potentially reduced capital cost relative to IEM-based devices, though losses in charge efficiency are incurred as a result of the diaphragm's lack of selectivity toward anions.  Hence, our results suggest that other types of binary electrolytes can be used with a diaphragm cation intercalation desalination device and that charge efficiency will depend on anion transference number.  Using porous electrode theory we show that similar desalination can be achieved relative to an IEM-based cell if high enough current is applied.  Based on observations from these simulations we derived closed-form equations that correlate the diaphragm's design and operating conditions to charge efficiency, and we subsequently use them to guide the selection of operating conditions for simulations at a range of current densities.  We quantify the tradeoffs between energy consumption and salt removal rate using accepted metrics for capacitive deionization technology and show that diaphragm-based cation intercalation desalination cells can be operated under highly energy-efficient conditions at slow rates, at high rates with low energy efficiency, or conditions in between depending on the applied current.  Based on an equilibrium theory and observations of simulated results, diffusion potential relations for the local degree of intercalation within electrodes are also derived, from which we find that intercalation processes are initially localized near the inlet of NRC cells as a result of streamwise salt concentration variations within electrodes.  In order to enhance reaction uniformity, we also propose a new architecture with a recycling electrolyte (i.e. the RC cell).  High flow rates in the RC cell enables ionic species to overcome ionic conductivity and concentration polarization limitations by producing more uniform reaction rates within porous electrodes. We also compare the cell potential curves and salt concentration distributions within diaphragm-based cells using flow-through, flow-behind, and flow-by configurations. Among these three the flow-through configuration shows the least variation of salt concentration within its electrode, while the other two configurations are more susceptible to the local depletion and accumulation of salt within their respective electrodes.  While the aforementioned results were obtained for Na$_{1+x}$NiFe(CN)$_6$ electrodes other cation intercalation electrodes could be used with a diaphragm.

Finally, we note that improvements to the present model can be made and that experimental demonstration has yet to be accomplished. This model does not incorporate the stability of the electroactive material in aqueous solution and side reactions. Also, the analytical model used to predict RC cell performance only produces accurate predictions when the tank volume is much larger than the electrode solution volume. In reality, the  mixture of various cationic species present in seawater and brackish water will affect the stability of the electroactive material, the functionality of the diaphragm, and the efficiency of ionic species transfer. Therefore, the current model could also be further advanced by investigating the effect of multiple charge/discharge cycles, multiple cations, and the hydraulic permeability of the diaphragm.  We also note that while our comparisons of energy consumption between diaphragm- and IEM-based cells show greater energy consumption for the diaphragm cell, our present IEM simulations are conservative in that the IEM-cell benchmark in this study is modeled as an infinitesimally thin interface with 100$\%$ anion-selectivity.  The finite resistance of real IEMs and the lack perfect anion selectivity will ultimately affect the comparison between these two architectures. 

\section{Acknowledgments}

The Department of Mechanical Science and Engineering at the University of Illinois at Urbana-Champaign supported this work. The numerical simulations conducted here made use of the aggregation-based algebraic multi-grid(AGMG) solver\cite{notay2015agmg,notay2010aggregation,notay2012aggregation,napov2012algebraic}.

\begin{appendices}
\label{sec:derivation}
\section{Analytical Salt Transport Model}
The analytical diaphragm salt-removal efficiency model presented here is based on several assumptions, which are pseudo-steady-state operation, negligible concentration variation along the direction perpendicular to the flow velocity, and negligible streamline-wise diffusion. For the NRC cell architecture, Eq.~\ref{e1} can be reduced to the following form for the positive electrode during charging:
\begin{equation}
    -\frac{dc_e}{dy}+\frac{t_{-}i}{u_{s}Fw_{e}}+\frac{2G_m^{\prime \prime}(c_{e,in}-c_{e})}{u_sw_e}=0\text{,}
    \label{e2}
\end{equation}
where $G_m^{\prime \prime}$ is the salt mass transfer conductance per unit cross-sectional area. $G_m^{\prime \prime}$ can be calculated as $G_m^{\prime \prime}=\frac{D_{eff,d}}{w_d+(\frac{\epsilon_d}{\epsilon_e})^{1.5}w_e}$ to include the diffusive salt transfer resistance in the electrode in addition to the diaphragm. By integrating Eq.(\ref{e2}) over the length of current collector $L_{cc}$, we obtain an expression for the concentration difference between the electrode's inlet and outlet ($\Delta c_e$):
\begin{equation}
    \Delta c_e = c_{e,out}^{+}-c_{e,in}=c_{e,in}-c_{e,out}^{-}=\Big\{1-\text{exp}\Big(-\frac{2L_{cc}G_m^{\prime \prime}}{u_{s}w_{e}}\Big)\Big\}\frac{t_{-}i}{2FG_m^{\prime \prime}}\text{.}
    \label{em2}
\end{equation}

For the RC cell, accepting the aforementioned assumptions, we further assume that salt solution mixes perfectly within each tank. The tank volume is set to be $V_{tank}$, and the equation for the salt accumulation rate in the tank connected to the positive electrode during charging is:
\begin{equation}
    \frac{dc_{tank}^+}{dt}=\frac{(c_{e,out}^{+}-c_{e,in}^{+})u_sw_eL_d}{V_{tank}}\text{.}
    \label{e4}
\end{equation}
The variables in this equation are also shown in Fig. \ref{fig:cell schematic}. Finally, we solve for the concentration difference $\Delta c_e$ between $c_e^0$ and $c_{e,tank}$ using Eqs. (\ref{e2}) and \ref{e4}:
\begin{equation}
    \Delta c_e = c_{tank}^+-c_{0}= c_{0}-c_{tank}^- = \frac{t_{-}i}{2FG_m^{\prime \prime}}\Big\{1-\text{exp}\Big(\frac{u_sw_eF\Delta c_{e,ideal}}{L_{cc}t_{-}i}\Big[\text{exp}\Big(-\frac{2L_{cc}G_m^{\prime \prime}}{w_eu_s}\Big)-1\Big]\Big)\Big\}\text{.} 
    \label{em3}
\end{equation}
Here, $\Delta c_{e,ideal}$ is $\Delta c_e$ for a cell with a perfect diaphragm. $\Delta c_{e,ideal}$ is calculated by integrating the salt accumulation rate expression of the tank (i.e., $\frac{dc}{dt}=\frac{L_dL_{cc}t_{-}i}{FV_{tank}}$) over the time required to finish the charging process. Thus, $\Delta c_{e,ideal}=\frac{L_dL_{cc}t_{-}w_e\nu_sc_{s,max}}{FV_{tank}}$, where $c_{s,max}$ is the terminal concentration of intercalated Na$^{+}$ in the IHC material in units of $\text{mol m}^{-3}$. $\Delta c_{e,ideal}$ from the ideal RC cell provides an upper limit of how much salt can be removed from source water.

\section{Diffusion Potential Equilibrium Relations}
The current density $\vec{i_e}$ in concentrated aqueous NaCl solution is expressed as \cite{smith2017theoretical} $\vec{i_e}=-\kappa_{eff}(\nabla{\phi_e}-\frac{2R_gT}{F}t_-\gamma_{\pm}\nabla{\ln{c_e}})$, where $\phi_e$, $t_{-}$, $\gamma_{\pm}$, $\kappa_{eff}$, and $R_gT/F$ are respectively solution-phase potential, anion transference number, the thermodynamic factor, the effective ionic conductivity, and the thermal potential. In the equilibrium limit where $\vec{i_e}\rightarrow0$ solution-phase potential gradients arise due to salt concentration gradients:

\begin{equation}
    \nabla \phi_{e}=\frac{2R_g Tt_-\gamma_{\pm}}{F}\nabla \ln{c_e}
    \label{B:2}
\end{equation}

\noindent Subsequently, we assume ideal solution thermodynamics such that $\gamma_{\pm}=1$, and we integrate Eq.~\ref{B:2} from point 1 to point 2 within a given porous electrode.  We thereby obtain the so-called ``diffusion potential'' difference between these two positions:

\begin{equation}
    \phi_{e,2}-\phi_{e,1}=\frac{2R_g Tt_-}{F} \ln{\bigg(\frac{c_{e,2}}{c_{e,1}}\bigg)}
    \label{B:3}
\end{equation}

\noindent We also assume that kinetic overpotential $\eta=\phi_s-\phi_e-\phi_{eq}$ is negligible for intercalation of cations, which is valid either when kinetics are facile or in an equilibrium setting. In either limit the equilibrium intercalation potential $\phi_{eq}$ at any position is equal to the difference between the solid-phase (i.e., electronic) potential $\phi_s$ and $\phi_e$ (i.e., $\phi_{eq}=\phi_s-\phi_e$).  If we further assume uniform distribution of $\phi_s$ in the IHC material, $\Delta \phi_{eq}$ is equivalent to $-\Delta \phi_e$ at two different points in electrolyte:

\begin{equation}
    \Delta\phi_{eq,12}=\phi_{eq,2}-\phi_{eq,1}=\phi_{e,1}-\phi_{e,2}=\frac{2R_g Tt_-}{F} \ln{\bigg(\frac{c_{e,1}}{c_{e,2}}\bigg)}
    \label{B:4}
\end{equation}

\noindent For a regular solution of cations and vacancies within NiHCF, as assumed in the present simulations and in Ref.~\cite{smith2017theoretical}, the equilibrium potential of intercalation varies with intercalated-Na fraction $x_{Na}$ as $\phi_{eq}=\phi_{eq}^0+\frac{R_g T}{F}\ln{(\frac{1-x_{Na}}{x_{Na}})}$, the relationship between the salt concentrations and the corresponding intercalated-Na fractions at both points can be established based on Eq.~\ref{B:4}:

\begin{equation}
    \Big(\frac{1-x_{Na,2}}{1-x_{Na,1}}\frac{x_{Na,1}}{x_{Na,2}}\Big)^{\frac{1}{2t_-}}=\frac{c_{e,1}}{c_{e,2}} \text{.}
\end{equation}

\noindent $x_{Na,2}$ can thereby be solved algebraically in terms of all other parameters: 

\begin{equation}
     x_{Na,2} = \frac{x_{Na,1}}{(\frac{c_{e,1}}{c_{e,2}})^{2t_-}-x_{Na,1}(\frac{c_{e,1}}{c_{e,2}})^{2t_-}+x_{Na,1}}\text{.}
\end{equation}
\noindent
\section{Supplementary information}
\subsection{Cycling parameters calculated from simulated data}

In all of the simulations cation intercalation desalination (CID) cells are subjected to one charge step and, subsequently, one discharge step to complete an electrochemical cycle.  Cells in either a non-recycling (NRC) or recycling (RC) configuration are simulated.  Here, a charge step corresponds to the application of current into the positive electrode, while a discharge step corresponds to the application of current away from it.  For each case cell potential and effluent salinity information are generated for the first cycle of operation using the spatiotemporal solutions obtained using porous electrode theory.  Here, cell potential is defined as the potential drop measured from the positive to the negative electrode, while effluent salinity is the salt concentration of desalinated water.  The following parameters are determined directly from the output of simulations:

1. $\Delta t_c$ is the time interval between two time steps during the charge step. Similarly, $\Delta t_d$ is an equivalent value during the discharge step.
\begin{equation}
    \Delta t = t_{i+1} - t_{i}
\end{equation}
where $i$ denotes the particular time step.

2. $t_c$ and $t_d$ represent the time elapsed during charge and discharge steps, respectively.
\begin{align}
\begin{split}
 t_c = \sum \Delta t_{c, i} ,
\\
 t_d = \sum \Delta t_{d, i} .
\end{split}
\end{align}

3. $c_{e,out}^c$ is the desalinated effluent salt concentration during the charging step. Its definitions for the different cells are:
\begin{equation}
    c_{e,out}^c = \sum_{i \text{ during charge}} \Delta t_{c,i} \times c_{e,i}^-/t_c \text{ for the NRC cell and}
\end{equation}
\begin{equation}
    c_{e,out}^c = c_{e,tank}^- \text{ for the RC cell,}
\end{equation}
 where $c_{e,i}^-$ is the effluent salt concentration of the negative electrode at each time step during a charge step, and $c_{e,tank}^-$ is the salt concentration in the tank connecting the negative electrode at the end of the charge step. Because $t_c$ is the time required to finish charge step, the time for the discharge step $t_d$ is used to calculate the effluent salt concentration during the discharge step $c_{e,out}^d$:
 \begin{equation}
    c_{e,out}^d = \sum_{i \text{ during discharge}} \Delta t_{d,i} \times c_{e,i}^+/t_d \text{ for the NRC cell, and}
\end{equation}
\begin{equation}
    c_{e,out}^d = c_{e,tank}^+ \text{ for the RC cell,}
\end{equation}
\noindent where $c_{e,tank}^+$ is the salt concentration in the tank connecting the positive electrode at the end of the discharge step.

5. $\bar V_{c}$ and $\bar V_{d}$ are the time averaged values of cell potential during the charging step and the discharging step, respectively:
\begin{equation}
    \bar V_{c} = \sum_{i \text{ during charge}} \Delta t_{c,i} \times V_{c,i}/t_c \text{ and}
\end{equation}
\begin{equation}
    \bar V_{d} = \sum_{i \text{ during discharge}} \Delta t_{d,i} \times V_{d,i}/t_c \text{ for both NRC and RC cell,}
\end{equation}
where $V_{d,i}$ and $V_{c,i}$ are the cell potentials at the end of time step ``i''.

\subsection{Metric definitions}

Each of the metrics used in this manuscript are defined as ratios of extensive quantities transferred during one cycle and geometric and gravimetric properties of the cycled cell.  The extensive quantities transferred include the moles of salt removed $N_{salt}$, the mass of salt removed $m_{salt}$, the volume of desalinated water produced $V_{desal}$, the net electrical energy consumed $E_{elec}$, and the moles of electrons transferred $N_{elec}$.  The metrics used in the manuscript include the desalination extent $\Delta c_e$, desalination energy consumption $E_d$, average salt adsorption rate (ASAR) expressed on the basis of electrode area $ASAR_a$ and IHC mass $ASAR_m$, energy normalized adsorbed salt (ENAS), and charge efficiency $\Gamma$. The definitions of these metrics are listed here:

1. $\Delta c_e$ is the ratio of the moles of salt removed to the volume of desalinated water: $\Delta c_e = N_{salt}/V_{desal}$.

2. $E_d$ is the ratio of the net electrical energy consumed to the volume of desalinated water: $E_d=E_{elec}/V_{desal}$.

3. $ASAR_a$ is the ratio of the moles of salt removed to the product of the projected area of one electrode $A$ and the time elapsed during the cycle $t_{cyc}$: $ASAR_a=N_{salt}/(A \times t_{cyc})$. $ASAR_m$ is the ratio of the mass of salt removed to the product of the mass of intercalation host compound (IHC) material in both electrodes $m_{IHC}$ and the time elapsed during the cycle $t_{cyc}$: $ASAR_m=m_{salt}/(m_{IHC} \times t_{cyc})$.

4. $ENAS$ is the ratio of the moles of salt removed to the net electrical energy consumed: $ENAS=N_{salt}/E_{elec}$.  Specific energy consumption is the inverse of $ENAS$.

5. $\it \Gamma$ is the ratio of the moles of salt removed to the moles of electrons transferred: $\it \Gamma=N_{salt}/N_{elec}$.  The diaphragm salt removal efficiency (DRSE) $\omega$ is equal to charge efficiency normalized by the anion transference number $t_-$: $\omega = \it \Gamma/t_-$.

\subsection{Metric expressions}

Using the fundamental definitions of each metric based on extensive quantities we now express the aforementioned metrics in terms of specific parameters calculated from cycling simulations.  In addition to the parameters described in Sec. 1 the material and cell parameters in Table 1 are also used.

\begin{table}[H]
\centering
\caption{Material and cell parameters used in the calculation metrics.}
\label{my-label}
\begin{tabular}{ll}
Parameter    & Definition                              \\
$i$             & current density (A m$^{-2}$)               \\
$z_+$           & cation valence charge number  \\
$F$           & Faraday's constant (C mol$^{-1}$)               \\
$t_-$         & anion transference number                \\
$u_s$         & superficial velocity (m s$^{-1}$)                \\
$L_{cc}$      & length of the current collector (m)       \\
$L_d$         & the depth of the cell into the page (m)  \\
$\upsilon_s$       & the volume fraction of IHC material in electrodes \\
$\rho_{IHC}$  & the density of IHC material (g cm$^{-3}$)            \\
$M_{salt}$    & molecular weight of salt (mg mol$^{-1}$)       \\
$V_{tank}$    & the volume of the tank (m$^{-3}$)           \\
\end{tabular}
\end{table}

\noindent For the desalination extent we have:
\begin{align}
\begin{split}
 \Delta c_e = \frac{t_d(c_0-c_{e,out}^d)+t_c(c_0-c_{e,out}^c)}{t_d + t_c} \text{ for the NRC cell and}
\\
 \Delta c_e = \frac{(c_0- c_{e,tank}^d)+(c_0-c_{e,tank}^c)}{2}\text{ for the RC cell}.
\end{split}
\end{align}

\noindent For the desalination energy consumption we have:
\begin{align}
\begin{split}
 E_d = \frac{iL_{cc}(t_c\bar V_c - t_d\bar V_d)}{u_sw_e(t_c+t_d)} \text{ for the NRC cell and}
\\
 E_d = \frac{iL_{cc}(t_c\bar V_c - t_d\bar V_d)}{2V_{tank}/L_d}\text{ for the RC cell}.
\end{split}
\end{align}

\noindent For the area-based average salt adsorption rate we have:
\begin{align}
\begin{split}
 ASAR_a = \frac{u_sw_e\Delta c_e}{L_{cc}} \text{ for the NRC cell and}
\\
 ASAR_a = \frac{2V_{tank}\Delta c_e}{L_{cc}L_d(t_d+t_c)}\text{ for the RC cell}.
\end{split}
\end{align}

\noindent For the mass-based average salt adsorption rate we have:

\begin{align}
\begin{split}
 ASAR_m = \frac{u_s\Delta c_eM_{salt}}{2L_{cc}\upsilon_s\rho_{IHC}} \text{ for the NRC cell and}
\\
 ASAR_m = \frac{V_{tank}M_{salt}\Delta c_e}{L_dL_{cc}w_e\upsilon_s\rho_{IHC}(t_d+t_c)}\text{ for the RC cell}.
\end{split}
\end{align}

\noindent For the charge efficiency we have:

\begin{align}
\begin{split}
 \it \Gamma = \frac{z_+F\Delta c_e u_sw_e}{iL_{cc}} \text{ for the NRC cell and}
\\
 \it \Gamma = \frac{2z_+FV_{tank}\Delta c_e}{iL_{cc}L_d(t_c+t_d)}\text{ for the RC cell}.
\end{split}
\end{align}

\noindent For the diaphragm salt removal efficiency we have:

\begin{align}
\begin{split}
 \omega = \frac{z_+F\Delta c_e u_sw_e}{it_-L_{cc}} \text{ for the NRC cell and}
\\
 \omega = \frac{2z_+FV_{tank}\Delta c_e}{it_-L_{cc}L_d(t_c+t_d)}\text{ for the RC cell}.
\end{split}
\end{align}

\end{appendices}
\clearpage
\medskip

\bibliographystyle{unsrt}


\end{document}